\pdfoutput=1

\documentclass[11pt,twoside,a4paper,cmspaper,final,collab]{cms-tdr}
\def\svnVersion{494742P}\def\cmsCernNoTag{CERN-EP-2018-305}\def\cmsCernDate{\today}\def\cmsMessage{Published in Physical Review D as \href{http://dx.doi.org/10.1103/PhysRevD.99.092005}{\doi{10.1103/PhysRevD.99.092005.}}}

\begin{document}\cmsNoteHeader{HIG-18-009}

\hyphenation{had-ron-i-za-tion}
\hyphenation{cal-or-i-me-ter}
\hyphenation{de-vices}
\RCS$HeadURL: svn+ssh://svn.cern.ch/reps/tdr2/papers/HIG-18-009/trunk/HIG-18-009.tex $
\RCS$Id: HIG-18-009.tex 493918 2019-04-16 13:44:56Z alverson $
\newlength\cmsFigWidth
\ifthenelse{\boolean{cms@external}}{\setlength\cmsFigWidth{0.49\textwidth}}{\setlength\cmsFigWidth{0.6\textwidth}}
\ifthenelse{\boolean{cms@external}}{\providecommand{\cmsLeft}{upper\xspace}}{\providecommand{\cmsLeft}{left\xspace}}
\ifthenelse{\boolean{cms@external}}{\providecommand{\cmsRight}{lower\xspace}}{\providecommand{\cmsRight}{right\xspace}}

\newcommand{\ZZ}{\ensuremath{\PZ\PZ}\xspace}
\newcommand{\WW}{\ensuremath{\PW\PW}\xspace}
\newcommand{\WZ}{\ensuremath{\PW\PZ}\xspace}
\newcommand{\tautau}{\ensuremath{\Pgt\Pgt}\xspace}
\newcommand{\ttW}{\ensuremath{\ttbar\PW}\xspace}
\newcommand{\ttZ}{\ensuremath{\ttbar\PZ}\xspace}
\newcommand{\ttH}{\ensuremath{\ttbar\PH}\xspace}
\newcommand{\ttG}{\ensuremath{\ttbar\gamma}\xspace}
\newcommand{\ttV}{\ensuremath{\ttbar\cmsSymbolFace{V}}\xspace}
\newcommand{\tZq}{\ensuremath{\cPqt\PZ\Pq}\xspace}
\newcommand{\tZW}{\ensuremath{\cPqt\PZ\PW}\xspace}
\newcommand{\tttt}{\ensuremath{\ttbar\ttbar}\xspace}
\newcommand{\WWqq}{\ensuremath{\PW^{\pm}\PW^{\pm}{\Pq\Pq}}\xspace}
\newcommand{\kappaV}{\ensuremath{\kappa_\text{V}}\xspace}
\newcommand{\kappat}{\ensuremath{\kappa_\cPqt}\xspace}
\newcommand{\ghvv}{\ensuremath{g_{\PH\mathrm{VV}}}\xspace}
\newcommand{\yt}{\ensuremath{y_\cPqt}\xspace}

\newcommand{\Hbb}{\ensuremath{\PH\to\bbbar}\xspace}
\newcommand{\HWW}{\ensuremath{\PH\to\PW\PW}\xspace}
\newcommand{\HZZ}{\ensuremath{\PH\to\PZ\PZ}\xspace}
\newcommand{\HTT}{\ensuremath{\PH\to\tautau}\xspace}
\newcommand{\tHq}{\ensuremath{\cPqt\PH\Pq}\xspace}
\newcommand{\tHW}{\ensuremath{\cPqt\PH\PW}\xspace}
\newcommand{\tH}{\ensuremath{\cPqt\PH}\xspace}
\newcommand{\tHttH}{\ensuremath{\cPqt\PH+\ttH}\xspace}

\newcommand{\epmup}{\ensuremath{\Pepm\Pgm^\pm}\xspace}
\newcommand{\mupmup}{\ensuremath{\Pgm^\pm\Pgm^\pm}\xspace}
\newcommand{\sstwol}{\ensuremath{2\ell\text{ss}}\xspace}
\newcommand{\threel}{\ensuremath{\ell\ell\ell}\xspace}
\newcommand{\gamgam}{\ensuremath{\gamma\gamma}\xspace}

\newcommand{\ttjets}{\ensuremath{\ttbar{+}\text{jets}}\xspace}
\newcommand{\Zjets}{\ensuremath{\PZ{+}\text{jets}}\xspace}
\newcommand{\tthf}{\ensuremath{\ttbar{+}\mathrm{HF}}\xspace}
\newcommand{\ttlf}{\ensuremath{\ttbar{+}\mathrm{LF}}\xspace}
\newcommand{\ttcc}{\ensuremath{\ttbar{+}\mathrm{c\bar{c}}}\xspace}
\newcommand{\ttbb}{\ensuremath{\ttbar{+}\bbbar}\xspace}
\newcommand{\ttb}{\ensuremath{\ttbar{+}\cPqb}\xspace}
\newcommand{\tttwob}{\ensuremath{\ttbar{+}2\cPqb}\xspace}
\newcommand{\thad}{\ensuremath{\cPqt_{\text{had}}}\xspace}
\newcommand{\Whad}{\ensuremath{\PW_{\text{had}}}\xspace}

\ifthenelse{\boolean{cms@external}}{\providecommand{\CL}{C.L.\xspace}}{\providecommand{\CL}{CL\xspace}}
\ifthenelse{\boolean{cms@external}}{\providecommand{\CLnp}{C.L\xspace}}{\providecommand{\CLnp}{CL}}
\newcolumntype{x}{>{\hfill}p{35pt}<{\hspace{5pt}}}
\newlength\cmsTabSkip\setlength{\cmsTabSkip}{1ex}

\cmsNoteHeader{HIG-18-009}
\title{Search for associated production of a Higgs boson and a single top quark in proton-proton collisions at \texorpdfstring{$\sqrt{s} = 13\TeV$}{sqrt(s) = 13 TeV}}

\date{\today}
\abstract{A search is presented for the production of a Higgs boson in association with a single top quark, based on data collected in 2016 by the CMS experiment at the LHC at a center-of-mass energy of $13\TeV$, which corresponds to an integrated luminosity of $35.9\fbinv$. The production cross section for this process is highly sensitive to the absolute values of the top quark Yukawa coupling, $\yt$, the Higgs boson coupling to vector bosons, $\ghvv$, and, uniquely, to their relative sign. Analyses using multilepton signatures, targeting $\PH\to\PW\PW$, $\PH\to\Pgt\Pgt$, and $\PH\to\PZ\PZ$ decay modes, and signatures with a single lepton and a $\bbbar$ pair, targeting the $\PH\to\bbbar$ decay, are combined with a reinterpretation of a measurement in the $\PH\to\gamgam$ channel to constrain $\yt$. For a standard model-like value of $\ghvv$, the data favor positive values of $\yt$ and exclude values of $\yt$ below about $-0.9\,\yt^\mathrm{SM}$.}

\hypersetup{
pdfauthor={CMS Collaboration},
pdftitle={Search for associated production of a Higgs boson and a single top quark in proton-proton collisions at sqrt(s) = 13 TeV},
pdfsubject={CMS},
pdfkeywords={CMS, higgs, top, physics}}

\maketitle

\section{Introduction}\label{sec:introduction}
The scalar resonance discovered by the CMS and ATLAS Collaborations at the LHC~\cite{Higgs-Discovery_ATLAS,Higgs-Discovery_CMS,Higgs-Discovery_CMS_long} in 2012 has been found to have properties consistent with the predictions of the standard model (SM) for a Higgs boson with a mass of about $125\GeV$~\cite{Aad:2015zhl}.
In particular, its couplings to bosons (\ghvv) and fermions ($y_\mathrm{f}$) corroborate an SM-like dependence on the respective masses.
Furthermore, data indicate that it has zero spin and positive parity~\cite{HIG-14-018}.
Recently, the associated production of top quark pairs with a Higgs boson (\ttH) and Higgs boson decays to pairs of bottom quarks have been observed~\cite{cms_tthobs,Aaboud:2018urx,HIG18016}, thereby directly probing the Yukawa interactions between the Higgs boson and top as well as bottom quarks for the first time.
In addition to measuring the absolute strengths of Higgs boson couplings, it is pertinent to assess the possible existence of relative phases among the couplings, as well as their general Lorentz structure.
Hence a broad sweep of Higgs boson production mechanisms and decay modes must be considered to reveal any potential deviations from the SM expectations.

The production rate of \ttH\ is sensitive only to the magnitude of the top quark-Higgs boson Yukawa coupling \yt\ and has no sensitivity to its sign.
Measurements of processes such as Higgs boson decays to photon pairs~\cite{Biswas:2012bd} or the associated production of \PZ\ and Higgs bosons via gluon-gluon fusion~\cite{Hespel:2015zea} on the other hand do have sensitivity to the sign, because of indirect effects in loop interactions.
Those measurements currently disfavor a negative value of the coupling~\cite{Khachatryan:2014jba,Khachatryan:2016vau}, but rely on the assumption that only SM particles contribute to the loops~\cite{Ellis}.

\begin{figure*}[!htb]
\centering
\includegraphics[width=0.30\textwidth]{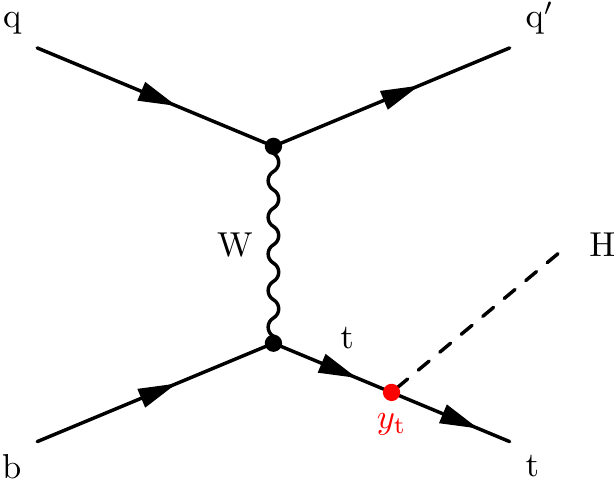} \hspace{\fill}
\includegraphics[width=0.30\textwidth]{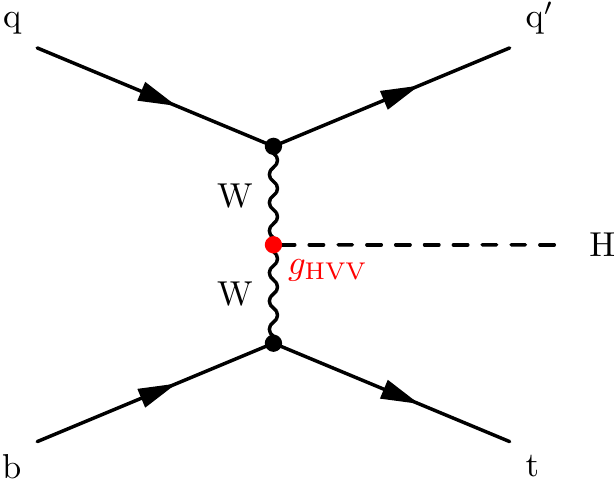} \hspace{\fill}
\includegraphics[width=0.30\textwidth]{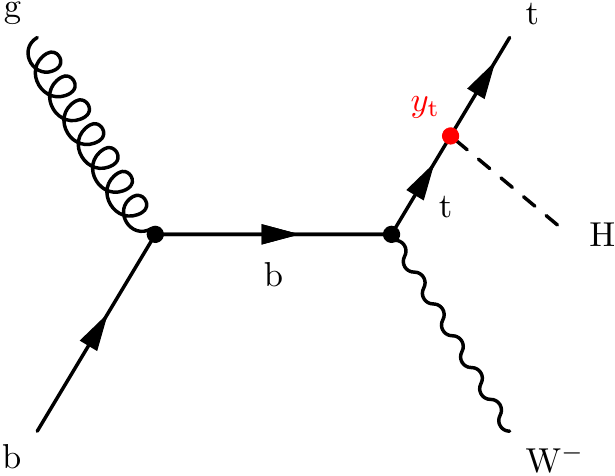}
\caption{Leading order Feynman diagrams for the associated production of a single top quark and a Higgs boson in the $t$ channel, where the Higgs boson couples either to the top quark (left) or the \PW\ boson (center), and one representative diagram of \tHW\ production, where the Higgs boson couples to the top quark (right).}
\label{fig:thqdiagrams}
\end{figure*}

In contrast, the production of Higgs bosons in association with single top quarks in proton-proton (\Pp\Pp) collisions proceeds via two categories of Feynman diagrams~\cite{Maltoni:2001hu,ennio,Agrawal:2012ga,Demartin:2015uha}, where the Higgs boson couples either to the top quark or the \PW\ boson.
The two leading order (LO) diagrams for the $t$ channel production process (\tHq) are shown in Fig.~\ref{fig:thqdiagrams}, together with one of the five LO diagrams for the \cPqt\PW\ process (\tHW), for illustration.
Because of the interference of these diagrams, the production cross section is uniquely sensitive to the magnitude as well as the relative sign and phase of the couplings.

In the SM, the interference between these two diagrams is maximally destructive and leads to very small production cross sections of about $71$, $16$, and $2.9$\unit{fb} for the $t$ channel, \cPqt\PW\ process, and $s$ channel, respectively, at a center-of-mass energy $\sqrt{s} = 13\TeV$~\cite{deFlorian:2016spz,Demartin:2016axk}.
Hence measurements using the data collected at the LHC so far are not yet sensitive to the SM production.
However, in the presence of new physics, there may be relative opposite signs between the \cPqt-\PH\ and \PW-\PH couplings which lead to constructive interference and enhance the cross sections by an order of magnitude or more.
In such scenarios, realized, \eg, in some two-Higgs doublet models~\cite{Celis:2013rcs}, \tHq\ production would exceed that of \ttH\ production, making it accessible with current LHC data sets.
In this paper, the \tHq\ and \tHW\ processes are collectively referred to as \tH\ production, while $s$ channel production is neglected.

The event topology of \tHq\ production is that of two heavy objects---the top quark, and the Higgs boson---in the central portion of the detector recoiling against one another, while a light-flavor quark and a soft \cPqb\ quark escape in the forward-backward regions of the detector.
Leptonic top quark decays produce high-momentum electrons and muons that can be used to trigger the detector readout.
Higgs boson decays to vector bosons or $\tau$ leptons ($\PH\to\PW\PW^*$, $\PZ\PZ^*$, or \tautau), which subsequently decay to light leptons ($\ell$ = $\Pepm$, $\Pgm^\pm$), lead to a multilepton final state with comparatively small background contributions from other processes.
Higgs boson decays to bottom quark-antiquark pairs ($\PH\to\bbbar$), on the other hand,  provide a larger event rate albeit with challenging backgrounds from \ttjets\ production.
In contrast, the rarer Higgs boson decays to two photons ($\PH\to\Pgg\Pgg$) result in easily triggered and relatively clean signals for both leptonic or fully hadronic top quark decays, with backgrounds mainly from other production modes of the Higgs boson.
The production of \tHW\ lacks the presence of forward activity and involves three heavy objects and therefore does not exhibit the defining features of \tHq\ events, while closely resembling the \ttH\ topologies, having identical final states.

The CMS Collaboration has previously searched for anomalous \tHq\ production in \Pp\Pp\ collision data at $\sqrt{s} = 8\TeV$, assuming a negative sign of the top quark Yukawa coupling relative to its SM value, $\yt = -\yt^{\mathrm{SM}}$, using all the relevant Higgs boson decay modes, and set limits on the cross section of this process~\cite{Khachatryan:2015ota}.
This paper describes two new analyses targeting multilepton final states and single-lepton + \bbbar\ final states, using a data set of \Pp\Pp\ collisions at $\sqrt{s}=13\TeV$ corresponding to an integrated luminosity of 35.9\fbinv, collected in 2016.
Furthermore, a previous measurement of Higgs boson properties in the $\PH\to\Pgg\Pgg$ final state at 13\TeV~\cite{Sirunyan:2018ouh} has been reinterpreted in the context of \tHq\ signal production and the results are included in a combination with those from the other channels.

This paper is structured as follows: the experimental setup and data samples are described in Sections~\ref{sec:experiment} and \ref{sec:datasimulation} respectively; the two analysis channels and their event selection, background estimations, and signal extraction techniques are described in Sections~\ref{sec:multilepton} and \ref{sec:bbchannels}; the reinterpretation of the $\PH\to\Pgg\Pgg$ result is described in Section~\ref{sec:aa}; and the results and interpretation are given in Section~\ref{sec:results}. The paper is summarized in Section~\ref{sec:conclusion}.

\section{The CMS experiment}\label{sec:experiment}
The central feature of the CMS apparatus is a superconducting solenoid of 6\unit{m} internal diameter, providing a magnetic field of 3.8\unit{T} along the beam direction.
Within the solenoid volume are a silicon pixel and strip tracker, a lead tungstate crystal electromagnetic calorimeter, and a brass and scintillator hadron calorimeter, each composed of a barrel and two endcap sections providing pseudorapidity coverage up to $\abs{\eta}<3.0$.
Forward calorimeters employing Cherenkov light detection extend the acceptance to  $\abs{\eta}<5.0$.
Muons are detected in gas-ionization chambers embedded in the steel flux-return yoke outside the solenoid with a fiducial coverage of $\abs{\eta}<2.4$.
The silicon tracker system measures charged particles within the range $\abs{\eta}<2.5$.
The impact parameters in the transverse and longitudinal direction are measured with an uncertainty of about 10 and 30\mum, respectively~\cite{Chatrchyan:2014fea}.
Tracks of isolated muons of transverse momentum $\pt\geq100\GeV$ and $\abs{\eta}<1.4$ are reconstructed with an efficiency close to 100\% and a $\pt$ resolution of about 1.3 to 2\% and smaller than 6\% for higher values of $\eta$.
For $\pt \leq 1 \TeV$ the resolution in the central region is better than 10\%.
A two-level trigger system is used to reduce the rate of recorded events to a level suitable for data acquisition and storage.
The first level of the CMS trigger system~\cite{Khachatryan:2016bia}, composed of custom hardware processors, uses information from the calorimeters and muon detectors to select the most interesting events in a time interval of less than 4\mus.
The high-level trigger processor farm further decreases the event rate from around 100\unit{kHz} to about 1\unit{kHz}.
A more detailed description of the CMS detector, together with a definition of the coordinate system and the kinematic variables used in the analysis, can be found in Ref.~\cite{Chatrchyan:2008aa}.

A full event reconstruction is performed by the particle-flow (PF) algorithm using optimized and combined information from all the subdetectors~\cite{Sirunyan:2017ulk}.
The individual PF candidates reconstructed are muons, electrons, photons, and charged and neutral hadrons, which are then used to reconstruct higher-level objects such as jets, hadronic taus, and missing transverse momentum (\ptmiss).
Additional quality criteria are applied to the objects to improve the selection purity.

Collision vertices are reconstructed using a deterministic annealing algorithm~\cite{Chabanat:2005zz,Fruhwirth:2007hz}.
The reconstructed vertex position is required to be compatible with the location of the LHC beam in the $x$--$y$ plane.
The vertex with the largest value of summed physics-object $\pt^2$ is considered to be the primary \Pp\Pp\ interaction (PV).
Charged particles, which are subsequently reconstructed, are required to be compatible with originating from the selected PV.

The identification of muons is based on linking track segments reconstructed in the silicon tracker and in the muon system~\cite{Chatrchyan:2012xi}.
If a link can be established, the track parameters are recomputed using the combination of hits in the inner and outer detectors.
Quality requirements are applied on the multiplicity of hits in the track segments, on the number of matched track segments, and on the quality of the track fit~\cite{Chatrchyan:2012xi}.

Electrons are reconstructed using an algorithm that matches tracks found in the silicon tracker with energy deposits in the electromagnetic calorimeter while limiting deposits in the hadronic calorimeter~\cite{Khachatryan:2015hwa}.
A dedicated algorithm takes into account the emission of bremsstrahlung photons and determines the energy loss~\cite{Khachatryan:2015iwa}.
A multivariate analysis (MVA) approach based on boosted decision trees (BDT) is employed to distinguish real electrons from hadrons mimicking an electron signature.
Additional requirements are applied in order to remove electrons originating from photon conversions~\cite{Khachatryan:2015hwa}.
Both muons and electrons from signal events are expected to be isolated, while those from heavy-flavor decays are often situated near jets.
Lepton isolation is quantified using the scalar \pt\ sum over PF candidates reconstructed within a cone centered on the lepton direction and shrinking with increasing lepton \pt.
The effect of additional \Pp\Pp\ interactions in the same and nearby bunch crossings (pileup) on the lepton isolation is mitigated by considering only charged particles consistent with the PV in the sum, and by subtracting an estimate of the contribution from neutral pileup particles within the cone area.

Jets are reconstructed from charged and neutral PF candidates using the anti-\kt\ algorithm~\cite{Cacciari:2008gp,Cacciari:2011ma} with a distance parameter of $0.4$, and with the constraint that the charged particles are compatible with the selected PV.
Jets originating from the hadronization of \cPqb\ quarks are identified using the ``combined secondary vertex'' (CSVv2) algorithm~\cite{Sirunyan:2017ezt}, which exploits observables related to the long lifetime of \cPqb\ hadrons and to the higher particle multiplicity and mass of \cPqb\ jets compared to light-quark and gluon jets.
Two working points of the CSVv2 discriminant output are used: a ``medium'' one, with a tagging efficiency for real \cPqb\ jets of $69\%$ and a probability of wrongly tagging jets from light-flavor quarks and gluons of about $1\%$, and a ``loose'' one, with a tagging efficiency of $83\%$ and a mistag rate for light-flavor jets of $8\%$.
Finally, the missing transverse momentum is defined as the magnitude of the vectorial \pt\ sum of all PF candidates in the event.

\section{Data and simulation}\label{sec:datasimulation}
Collision events for this analysis are selected by the following high-level trigger algorithms.
Events in the multilepton channels must pass at least one of single-lepton, dilepton, or trilepton triggers with loose identification and isolation requirements and with a minimum \pt\ threshold based on the lepton multiplicity in the final state.
Events in the single lepton + \bbbar\ channels must pass the same single-lepton triggers, or a dilepton trigger for the control region described in Section~\ref{sec:bbchannels}.
The minimum $\pt$ threshold for single lepton triggers is 24\GeV for muons and 27\GeV for electrons.
For dilepton triggers, the $\pt$ thresholds on the leading and subleading leptons are 17\GeV and 8\GeV for muons, and 23\GeV and 12\GeV for electrons, respectively.
For the trilepton trigger, the third hardest lepton $\pt$ must be greater than 5\GeV for muons and 9\GeV for electrons.

The data are compared to signal and background estimations based on Monte Carlo (MC) simulated samples and techniques based on control samples in data.
All simulated samples include the response of the CMS detector based on the \GEANTfour~\cite{GEANT} toolkit and are generated with a Higgs boson mass of 125\GeV and a top quark mass of 172.5\GeV.
The event generator used for the \tHq\ and \tHW\ signal samples is \MGvATNLO\ (version 2.2.2)~\cite{amcatnlo} at LO precision~\cite{Alwall:2007fs} and using the \textsc{NNPDF3.0} set of parton distribution functions (PDF)~\cite{Ball:2014uwa} with the \textsc{PDF4LHC} prescription~\cite{Botje:2011sn,Alekhin:2011sk}.
The samples are normalized to next-to-leading order (NLO) SM cross sections at 13\TeV of 71.0 and 15.6\unit{fb} for \tHq\ and \tHW, respectively~\cite{deFlorian:2016spz,Demartin:2016axk}.

The Higgs boson production cross sections and branching fractions are expressed as functions of Higgs boson coupling modifiers in the kappa framework~\cite{Heinemeyer:2013tqa}, where the coupling modifiers $\kappa$ are defined as the ratio of the actual value of a given coupling to the one predicted by the SM.
Particularly relevant for the \tH\ case are the top quark and vector boson coupling modifiers: $\kappat\equiv\yt/\yt^{\mathrm{SM}}$ and $\kappaV\equiv\ghvv/\ghvv^{\mathrm{SM}}$, where V stands for either \PW\ or \PZ\ bosons.
The dependence of the \tHq\ and \tHW\ production cross sections on \kappat\ and \kappaV\ are assumed to be as follows (calculated at NLO using \MGvATNLO~\cite{deFlorian:2016spz,Demartin:2015uha,Demartin:2016axk}):
\begin{align*}
	\sigma_{\tHq} &= (2.63\,\kappat^2 + 3.58\,\kappaV^2 - 5.21\,\kappat\kappaV) \sigma_{\tHq}^{\mathrm{SM}}, \\
	\sigma_{\tHW} &= (2.91\,\kappat^2 + 2.31\,\kappaV^2 - 4.22\,\kappat\kappaV) \sigma_{\tHW}^{\mathrm{SM}}.
\end{align*}
Event weights are produced in the generation of both samples corresponding to $33$ values of \kappat\ between $-6.0$ and $+6.0$, and for $\kappaV=1.0$.
The \tHq\ events are generated with the four-flavor scheme (4FS) while the \tHW\ process uses the five-flavor scheme (5FS) to disentangle the LO interference with the \ttH\ process~\cite{Demartin:2016axk}.

The \MGvATNLO\ generator is also used for simulation of the \ttH\ process and the main backgrounds: associated production of \ttbar\ pairs with vector bosons, \ttW, \ttZ, at NLO~\cite{Frederix:2012ps}, and with additional jets or photons, \ttjets, $\ttbar\gamma+\text{jets}$ at LO.
All the rates are normalized to next-to-next-leading order cross sections.
In particular, the \ttH\ production cross section is taken as 0.507\unit{pb}~\cite{deFlorian:2016spz}.
A set of minor backgrounds are also simulated with \MGvATNLO\ at LO, or with other generators, such as NLO \POWHEG v2~\cite{Nason:2004rx,Frixione:2007vw,Alioli:2010xd,Re:2010bp,Alioli:2009je,Melia:2011tj}.
All generated events are interfaced with \PYTHIA\ (8.205)~\cite{PYTHIA8} for the parton shower and hadronization steps.

The object reconstruction in MC events uses the same algorithm as used in data.
Furthermore, the trigger selection is simulated and applied for generated signal events.
However, the triggering and selection efficiencies for leptons are different between data and simulation, at the level of 1$\%$.
All simulated events used in the analyses are corrected by applying small data-to-MC scale factors to improve the modeling of the data.
Separate scale factors are applied to correct for the difference in trigger efficiency, lepton reconstruction and selection efficiency, as well as the \cPqb\ tagging efficiency and the resolution of the missing transverse momentum.

Simulated events are weighted according to the number of pileup interactions so that the distribution of additional \Pp\Pp\ interactions in the simulated samples matches that observed in data, as estimated from the measured bunch-to-bunch instantaneous luminosity and the total inelastic cross section.

\section{Multilepton channels}\label{sec:multilepton}
Signal \tH\ events where the top quark decay produces leptons and the Higgs boson decays to vector bosons or \Pgt\ leptons can lead to final states containing multiple isolated, high-\pt\ leptons with different charge and flavor configurations.
Of particular interest among these are those with three or more charged leptons or with two leptons of the same electric charge, as they appear with comparatively low backgrounds.
Selecting such events in $\Pp\Pp$ collisions while requiring the presence of \cPqb-tagged jets typically yields a mixture of mostly \ttjets\ events with nonprompt leptons and events from the associated production of \ttbar\ with a vector boson (\ttW\ and \ttZ) or with a Higgs boson (\ttH) that decay to additional prompt leptons.
The analysis described in this section separates the \tHq\ signal from these two dominant background sources by training two multivariate classifiers using features such as the forward light jet, the difference in multiplicities of jets and \cPqb-tagged jets (``\cPqb\ jets''), as well as the kinematic properties of the leptons.
The two classifier outputs are combined into a single binned distribution, which is then fit to the data to extract the signal yield and constrain the background contributions.

\subsection{Event and object selections}\label{ssec:multilepselection}
In the multilepton channels, events are selected with trigger algorithms involving one, two, or three leptons passing the given \pt\ thresholds.
At the offline analysis level, a distinction is made between prompt signal leptons (from \PW, \PZ, or leptonic \Pgt\ decays) and nonprompt leptons (either genuine leptons from heavy-flavor hadron decays, asymmetric \Pgg\ conversions, or jets misidentified as leptons).
For this purpose an MVA classifier is used~\cite{tthlepton}, exploiting the properties of the jet associated with individual leptons in addition to the lepton kinematics, isolation, and reconstruction quality.
The leptons are selected if they pass a certain threshold of the classifier output and are referred to as ``tight'' leptons, with a lower threshold defined for a relaxed selection and ``loose'' leptons.

The final \tH\ event selection targets signatures with \HWW\ and $\cPqt\to\PW\cPqb\to \ell\nu\cPqb$, which results in three \PW\ bosons, one \cPqb\ quark, and a light quark at high rapidity.
Three mutually exclusive channels are defined based on the number of tight leptons and their flavors: exactly two same-sign leptons (\sstwol), either \mupmup\ or \epmup, or exactly three leptons (\threel, $\ell=\Pgm$ or $\Pe$).
The same-sign dielectron channel suffers from larger backgrounds and does not add sensitivity and is therefore not included in the analysis.
There is an additional requirement of at least one \cPqb-tagged jet (using the medium working point of the CSVv2 algorithm) and at least one light-flavor (untagged, using the loose working point) jet.
The full selection is summarized in Table~\ref{tab:mlcuts}.

\begin{table}[!ht]
  \topcaption{Summary of the event selection for the multilepton channels.\label{tab:mlcuts}}
  \centering
    \begin{scotch}{p{8cm}}
        \quad Same-sign channel (\mupmup or \epmup)  \\
        Exactly two tight SS leptons            \\
        $\pt>25/15\GeV$                         \\
        No loose leptons with $m_{\ell\ell} < 12\GeV$ \\
        One or more \cPqb-tagged jet with $\pt>25\GeV$ and $\abs{\eta}<2.4$ \\
        One or more untagged jets with $\pt>25\GeV$ for $\abs{\eta}<2.4$ and $\pt>40\GeV$ for $\abs{\eta}>2.4$ \\
        [\cmsTabSkip]
 		\quad $\threel$ channel \\
        Exactly three tight leptons               \\
        $\pt>25/15/15\GeV$               \\
        No lepton pair with $\abs{m_{\ell\ell}-m_\PZ}<15\GeV$ \\
        No loose leptons with $m_{\ell\ell} < 12\GeV$ \\
        One or more \cPqb-tagged jet with $\pt>25\GeV$ and $\abs{\eta}<2.4$ \\
        One or more untagged jets with $\pt>25\GeV$ for $\abs{\eta}<2.4$ and $\pt>40\GeV$ for $\abs{\eta}>2.4$ \\
    \end{scotch}
\end{table}

About one quarter of the events in the finally selected sample are from \HTT\ and \HZZ\ decays, with the rest coming from \HWW\ decays, as determined from the \tHq\ signal simulation.
A significant fraction of selected events also pass the selection used in the dedicated search for \ttH\ in multilepton channels~\cite{tthlepton}: about 50\% in the dilepton channels and about 80\% in the \threel\ channels.

\subsection{Backgrounds}\label{ssec:multilepbackgrounds}
The background processes contributing to the signal sample can be divided into two classes, reducible and irreducible, and are estimated respectively from data and MC simulation.
Irreducible physics processes, such as the associated production of an electroweak boson with a top quark pair (\ttV, $\mathrm{V}=\PW, \PZ$), give rise to final states very similar to the \tHq\ signal and are directly estimated from MC simulation.
However, the dominant contribution is from the reducible background arising from nonprompt leptons, mainly from \ttbar\ production.
This background is suppressed to a certain extent by tightening the lepton selection criteria.
The background estimation methods employed here and summarized below are identical to those used in the dedicated search for \ttH\ in multilepton channels~\cite{tthlepton}.

The yield of reducible backgrounds is estimated from the data, using a ``tight-to-loose'' ratio measured in a control region dominated by nonprompt leptons.
The ratio represents the probability with which the nonprompt leptons that pass the looser selection can also pass the tight criteria, and is measured in categories of the lepton \pt\ and $\eta$.
A sideband region in data which has loosely selected leptons is then extrapolated with this ratio to obtain the nonprompt background contribution.

A further background in the same-sign dilepton channels arises from events where the charge of one lepton is wrongly assigned.
This can be estimated from the data, by measuring the charge misidentification probability using the \PZ\ boson mass peak in same-sign dilepton events, and weighting events with opposite-sign leptons to determine the yield in the signal region.
The effect is found to be negligible for muons but sizable for electrons.

The production of \WZ\ pairs with leptonic \PZ\ boson decays has similar leptonic features as the signal, but usually lacks the hadronic activity required in the signal selection.
To determine the corresponding diboson contribution in the signal region, simulated \WZ\ events have been used along with a normalization scale factor determined from data in an exclusive control region.

Other subdominant backgrounds are estimated from MC simulation and include additional multiboson production, such as \ZZ, \WWqq, VVV, same-sign \PW\ boson production from double-parton scattering (DPS), associated production of top quarks with \PZ\ bosons (\tZq, \tZW), events with four top quarks, and \ttbar\ production in association with photons and subsequent asymmetric conversions.

The expected and observed event yields after the selections described in Table~\ref{tab:mlcuts} are shown in Table~\ref{tab:mlyields}.

\begin{table*}[!htbp]
  \topcaption{Data yields and expected backgrounds after the event selection for the three multilepton search channels in 35.9\fbinv\ of integrated luminosity. Quoted uncertainties include statistical uncertainties reflecting the limited size of MC samples and data sidebands, and unconstrained systematic uncertainties.\label{tab:mlyields}}
  \centering
  \begin{scotch}{lrrr}
  Process & \mupmup & \epmup  & $\ell\ell\ell$ \\
  \hline
  $\ttW$                        & $ 68   \pm 10   $ & $97     \pm 13   $ & $22.5   \pm  3.1 $ \\
  $\ttZ/\ttG$                   & $ 25.9 \pm  3.9 $ & $64.8   \pm  9.0 $ & $32.8   \pm  5.1 $ \\
  $\WZ$                         & $ 15.1 \pm  7.7 $ & $26     \pm 13   $ & $ 8.2   \pm  2.4 $ \\
  $\ZZ$                         & $  1.16\pm  0.65$ & $2.9    \pm  1.5 $ & $ 1.62  \pm  0.87$ \\
  $\WWqq$                       & $  4.0 \pm  2.1 $ & $7.0    \pm  3.6 $ & \NA\               \\
  $\PW^\pm\PW^\pm$ (DPS)        & $  2.5 \pm  1.3 $ & $4.2    \pm  2.2 $ & \NA\               \\
  VVV                           & $  3.0 \pm  1.5 $ & $4.9    \pm  2.5 $ & $ 0.42  \pm  0.26$ \\
  $\tttt$                       & $  2.3 \pm  1.2 $ & $4.1    \pm  2.1 $ & $ 1.8   \pm  1.0 $ \\
  $\tZq$                        & $  5.8 \pm  3.6 $ & $10.7   \pm  6.1 $ & $ 3.9   \pm  2.5 $ \\
  $\tZW$                        & $  2.1 \pm  1.1 $ & $3.9    \pm  2.0 $ & $ 1.70  \pm  0.86$ \\
  $\gamma$ conversions          & \NA\              & $23.8   \pm  7.8 $ & $ 7.4   \pm  2.8 $ \\ [\cmsTabSkip]
  Nonprompt                     & $ 80.9 \pm  9.4$  & $135    \pm 35   $ & $26     \pm 14   $ \\
  Charge misidentification      & \NA\              & $58     \pm 17   $ & \NA\               \\ [\cmsTabSkip]
  Total background              & $211   \pm 17   $ & $443    \pm 45   $ & $106    \pm 16   $ \\ [\cmsTabSkip]
  $\ttH$                        & $ 24.2 \pm  2.1 $ & $ 35.2  \pm  2.9 $ & $ 18.3  \pm  1.7 $ \\
  $\tHq$ (SM)                   & $  1.43\pm  0.12$ & $  1.92 \pm  0.15$ & $  0.52 \pm  0.04$ \\
  $\tHW$ (SM)                   & $  0.71\pm  0.06$ & $  1.11 \pm  0.09$ & $  0.62 \pm  0.05$ \\ [\cmsTabSkip]
  Total SM                      & $237   \pm 17   $ & $482    \pm 45   $ & $126    \pm 16   $ \\ [\cmsTabSkip]
  $\tHq$ ($\kappaV=1=-\kappat$) & $ 18.5 \pm  1.6 $ & $ 27.4  \pm  2.1 $ & $  7.48 \pm  0.58$ \\
  $\tHW$ ($\kappaV=1=-\kappat$) & $  7.72\pm  0.65$ & $ 11.23 \pm  0.91$ & $  7.38 \pm  0.60$ \\ [\cmsTabSkip]
  Data                          &  280              &  525               &  127               \\
  \end{scotch}
\end{table*}

\subsection{Signal extraction}\label{ssec:multilepsignalextrac}
After applying the event selection of the multilepton channels, only about one percent of selected events are expected to be from \tH\ production (assuming SM cross sections), while roughly 10\% of events are from \ttH\ production.
To discriminate this small signal from the backgrounds, an MVA method is employed: a classification algorithm is trained twice with \tHq\ events as the signal class, and either \ttV\ (mixing \ttW\ and \ttZ\ according to their respective cross sections) or \ttjets\ as background classes.
The two separate trainings allow the exploitation of the different jet and \cPqb\ jet multiplicity distributions, and of the different kinematic properties of the leptons in the two dominant background classes.
Several machine learning algorithms were studied for potential use, and the best performance was obtained with a gradient BDT using a maximum tree depth of three and an ensemble of 800 trees~\cite{Hocker:2007ht}.
Events from \tHW\ and \ttH\ production are not used in the training and, because of their kinematic similarity with the \ttV\ background, tend to be classified as backgrounds.

As observed above, the features of the \tHq\ signal can be split into three broad categories: those related to the forward jet activity; those related to jet and \cPqb-jet multiplicities; and those related to kinematic properties of leptons, as well as their total charge.
A set of ten observables were used as input features to the classification training, and are listed in Table~\ref{tab:mlbdtinputs}.
The training is performed separately for the \sstwol\ and the \threel\ channels with the same or equivalent input features.

\begin{table}[ht!]
\topcaption{Input observables to the signal discrimination classifier.}
\centering
\begin{scotch}{lp{12cm}}
& Number of jets with $\pt>25\GeV$, $\abs{\eta}<2.4$\\
& Maximum $\abs{\eta}$ of any (untagged) jet (``forward jet'')\\
& Sum of lepton charges \\
& Number of untagged jets with $\abs{\eta}>1.0$\\
& $\Delta \eta$ between forward light jet and leading \cPqb-tagged jet\\
& $\Delta \eta$ between forward light jet and subleading \cPqb-tagged jet \\
& $\Delta \eta$ between forward light jet and closest lepton\\
& $\Delta \phi$ of highest-\pt\ same-sign lepton pair\\
& Minimum $\Delta R$ between any two leptons\\
& \pt\ of subleading (or $3^{rd}$) lepton\\
\end{scotch}
\label{tab:mlbdtinputs}
\end{table}

\begin{figure*}[htb]
  \centering
    \includegraphics[width=0.32\linewidth]{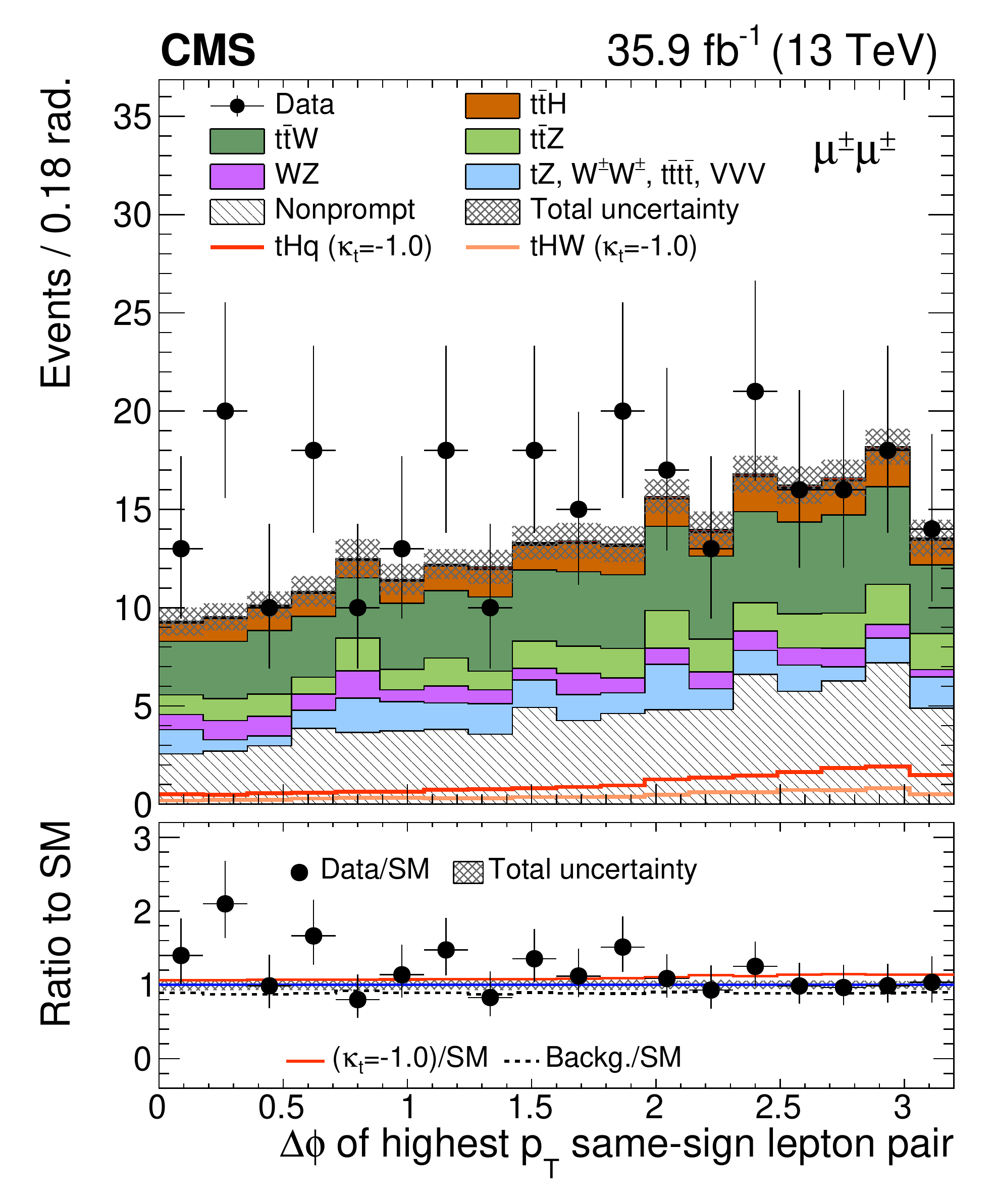}
    \includegraphics[width=0.32\linewidth]{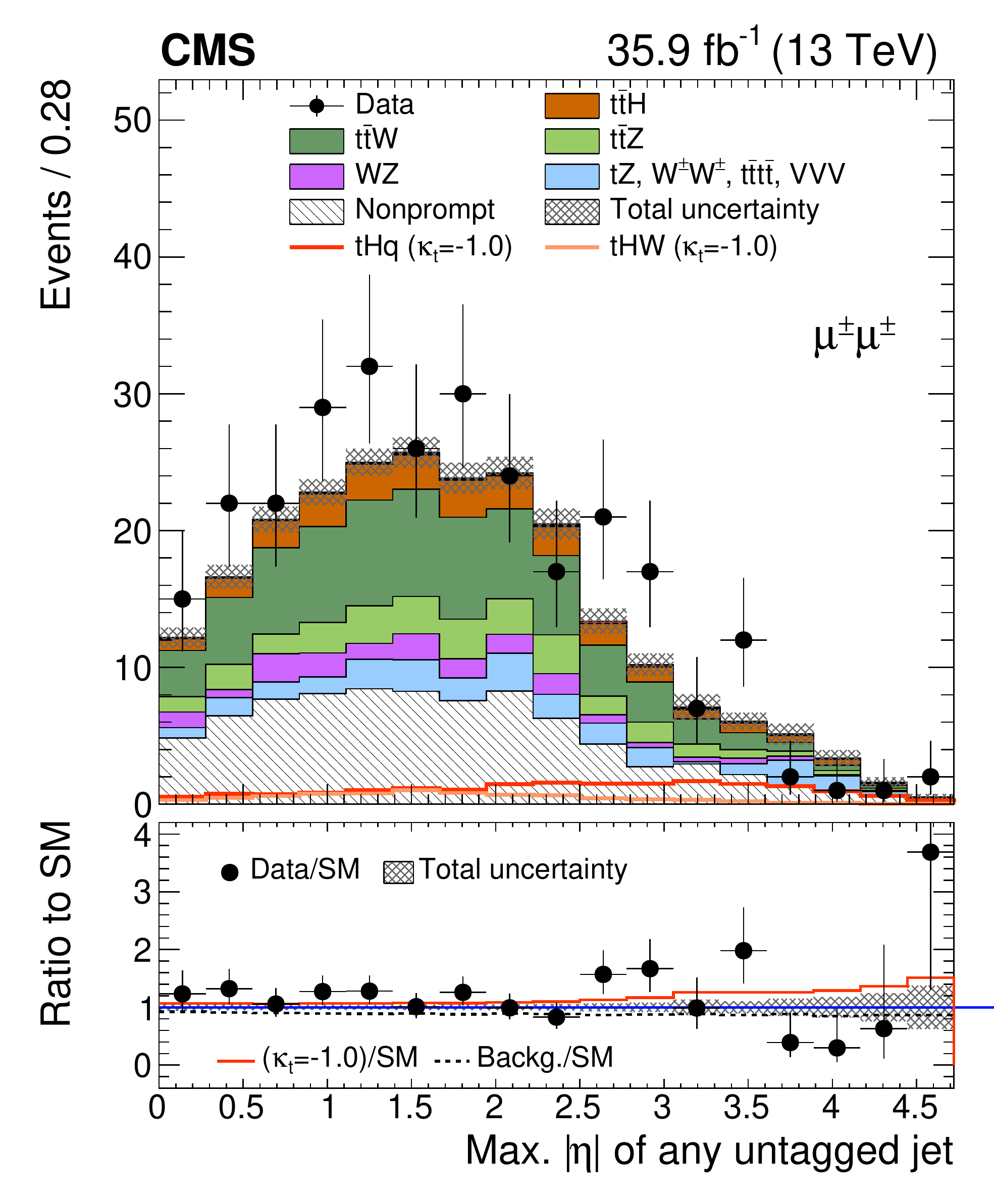}
    \includegraphics[width=0.32\linewidth]{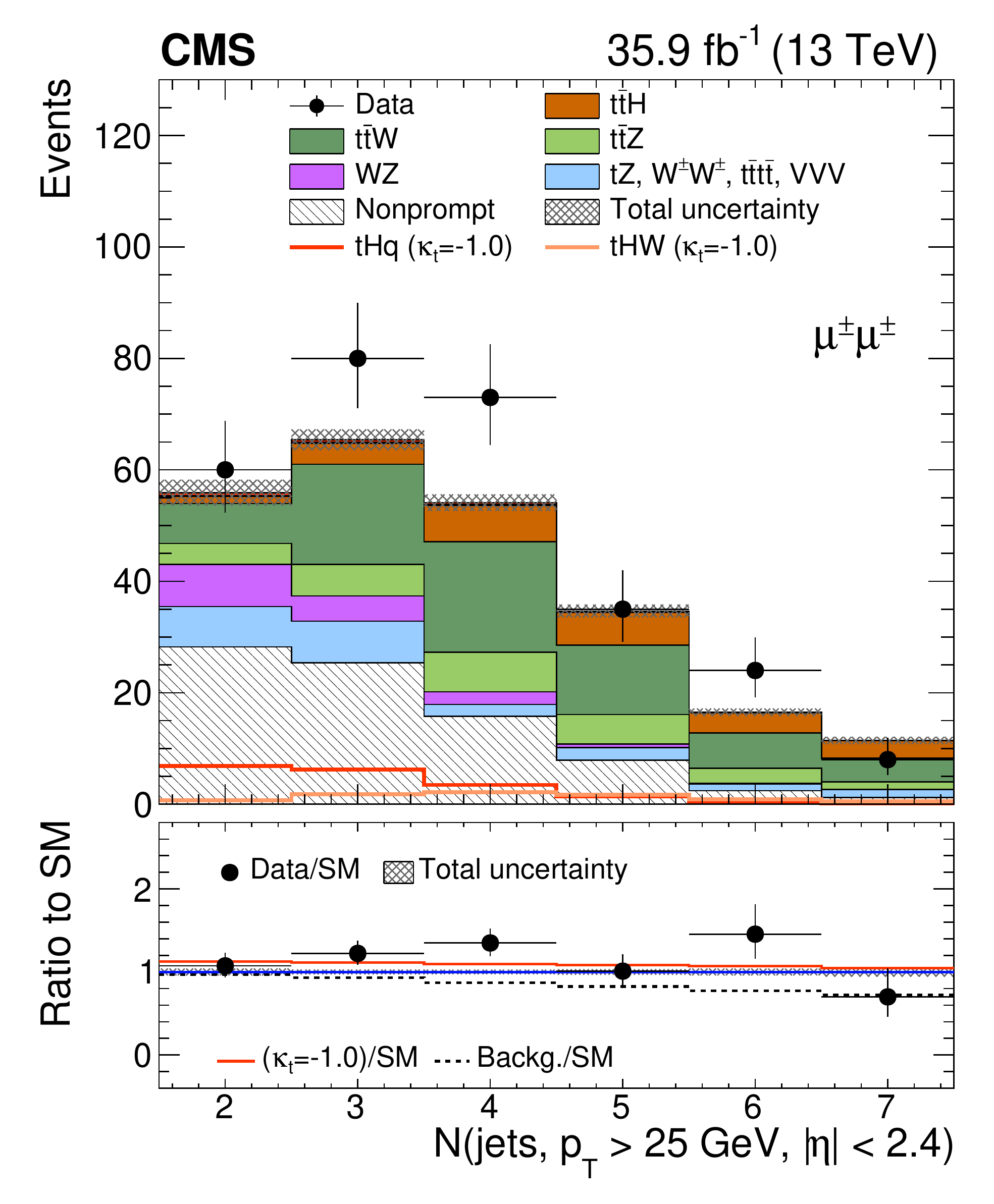} \\
  \caption{Distributions of discriminating observables for the same-sign \mupmup\ channel, normalized to 35.9\fbinv, before fitting the signal discriminant to the data. The grey band represents the unconstrained (pre-fit) statistical and systematic uncertainties. In the panel below each distribution, the ratio of the observed and predicted event yields is shown. The shape of the two \tH\ signals for $\kappat=-1.0$ is shown, normalized to their respective cross sections for $\kappat=-1.0, \kappaV=1.0$.\label{fig:2lss_inputs_mm}}
\end{figure*}

\begin{figure*}[htb]
  \centering
    \includegraphics[width=0.32\linewidth]{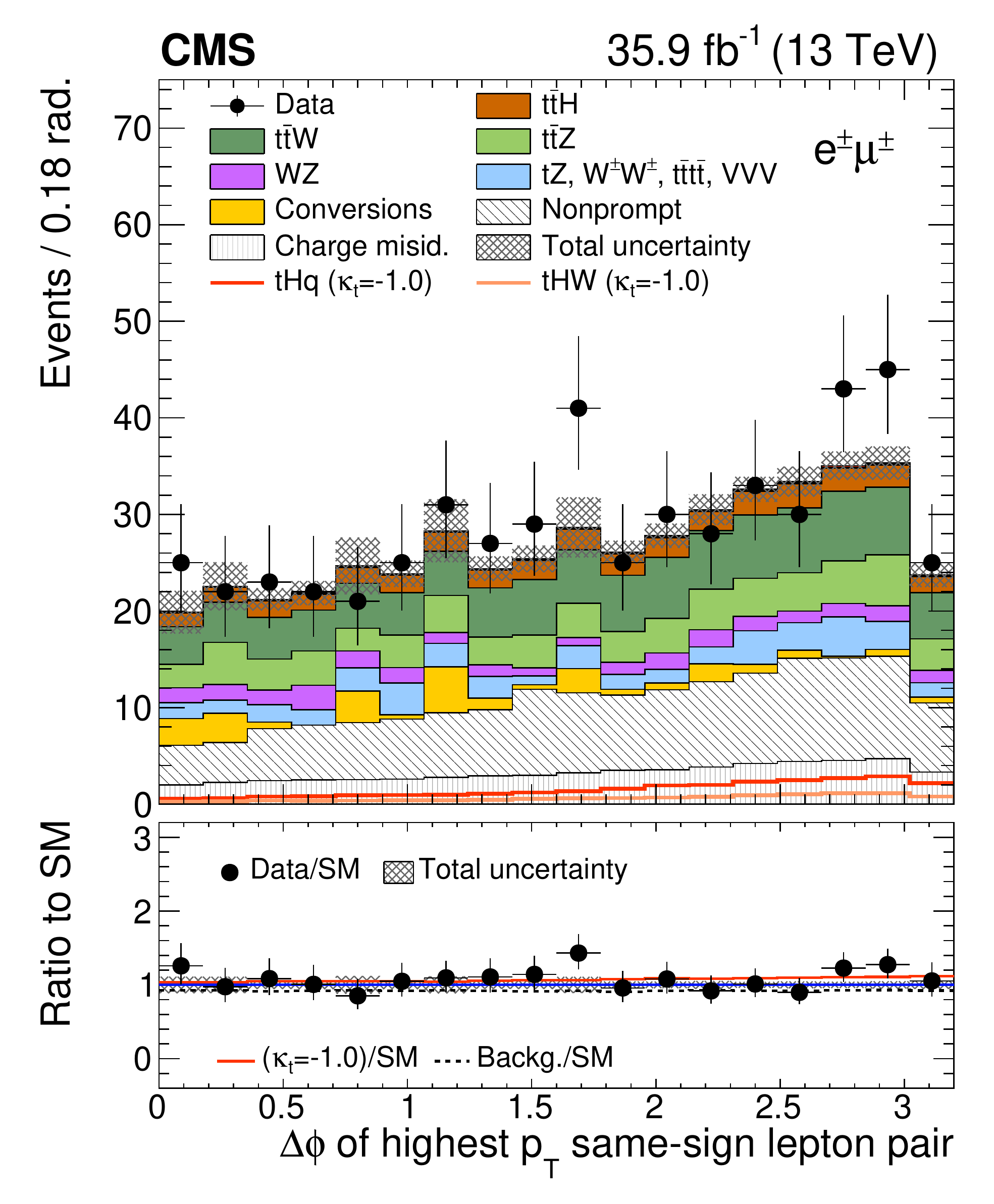}
    \includegraphics[width=0.32\linewidth]{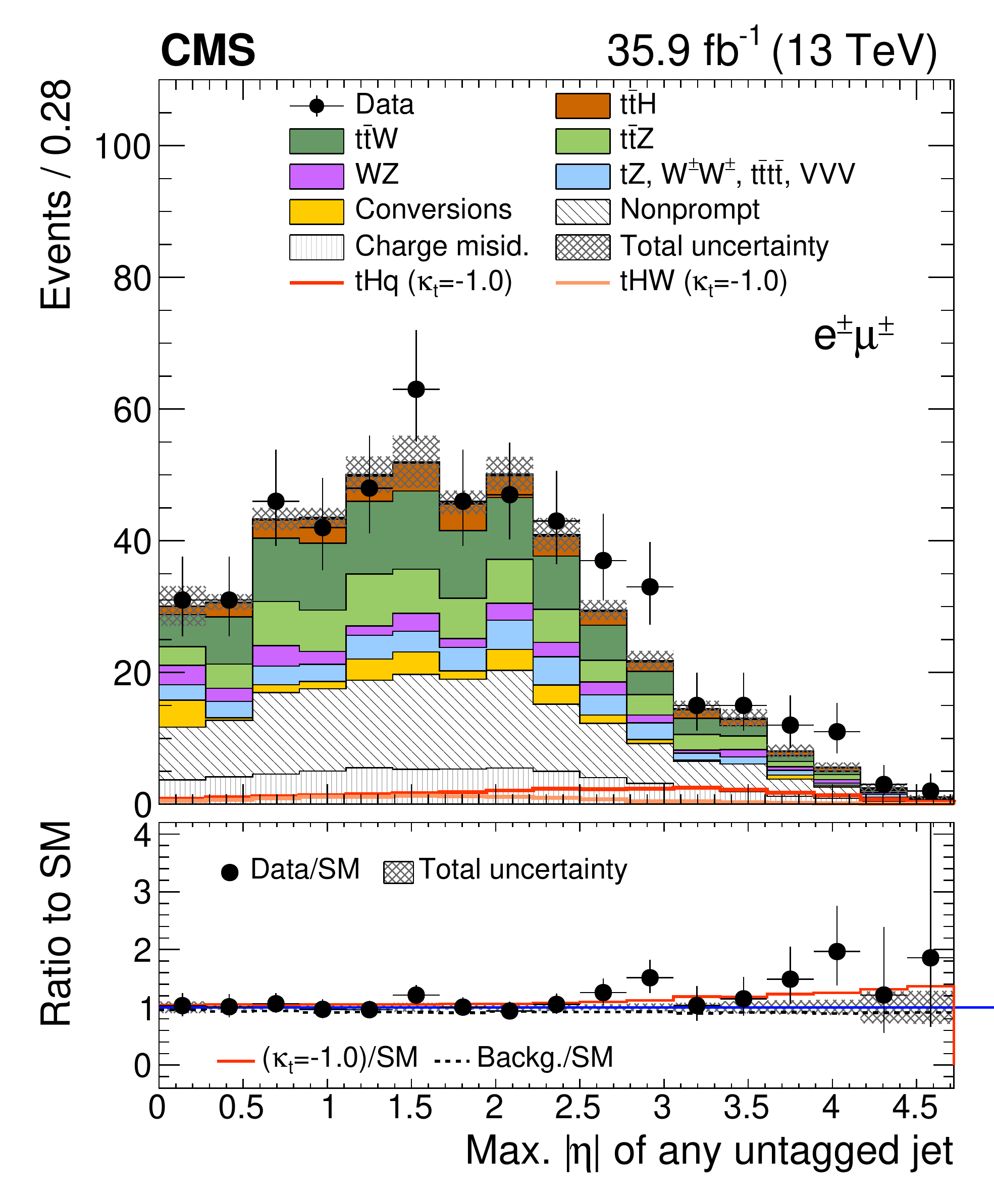}
    \includegraphics[width=0.32\linewidth]{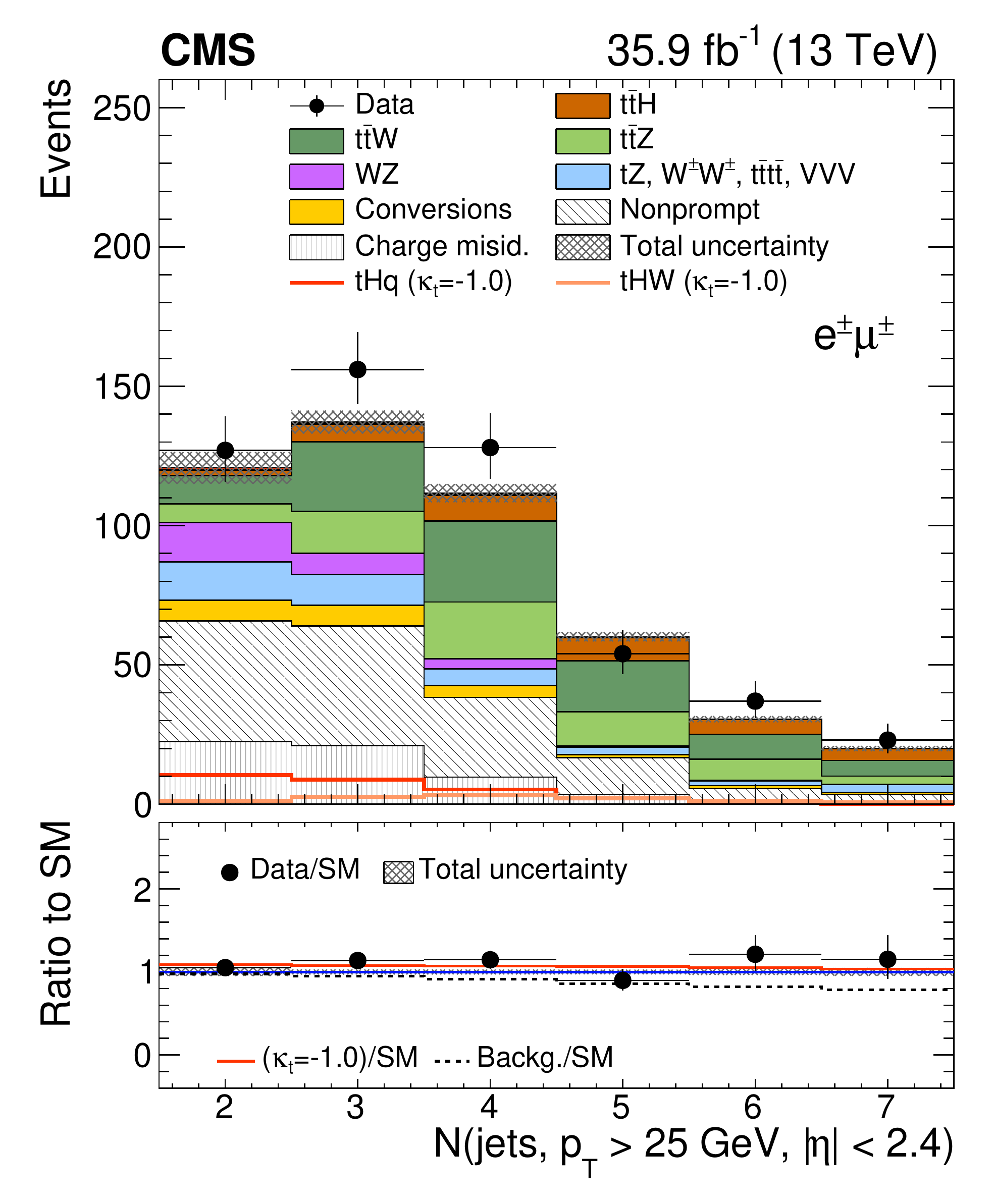} \\
  \caption{Distributions of discriminating observables for the same-sign \epmup\ channel, normalized to 35.9\fbinv, before fitting the signal discriminant to the data. The grey band represents the unconstrained (pre-fit) statistical and systematic uncertainties. In the panel below each distribution, the ratio of the observed and predicted event yields is shown. The shape of the two \tH\ signals for $\kappat=-1.0$ is shown, normalized to their respective cross sections for $\kappat=-1.0, \kappaV=1.0$.\label{fig:2lss_inputs_em}}
\end{figure*}

\begin{figure*}[htb]
  \centering
    \includegraphics[width=0.32\linewidth]{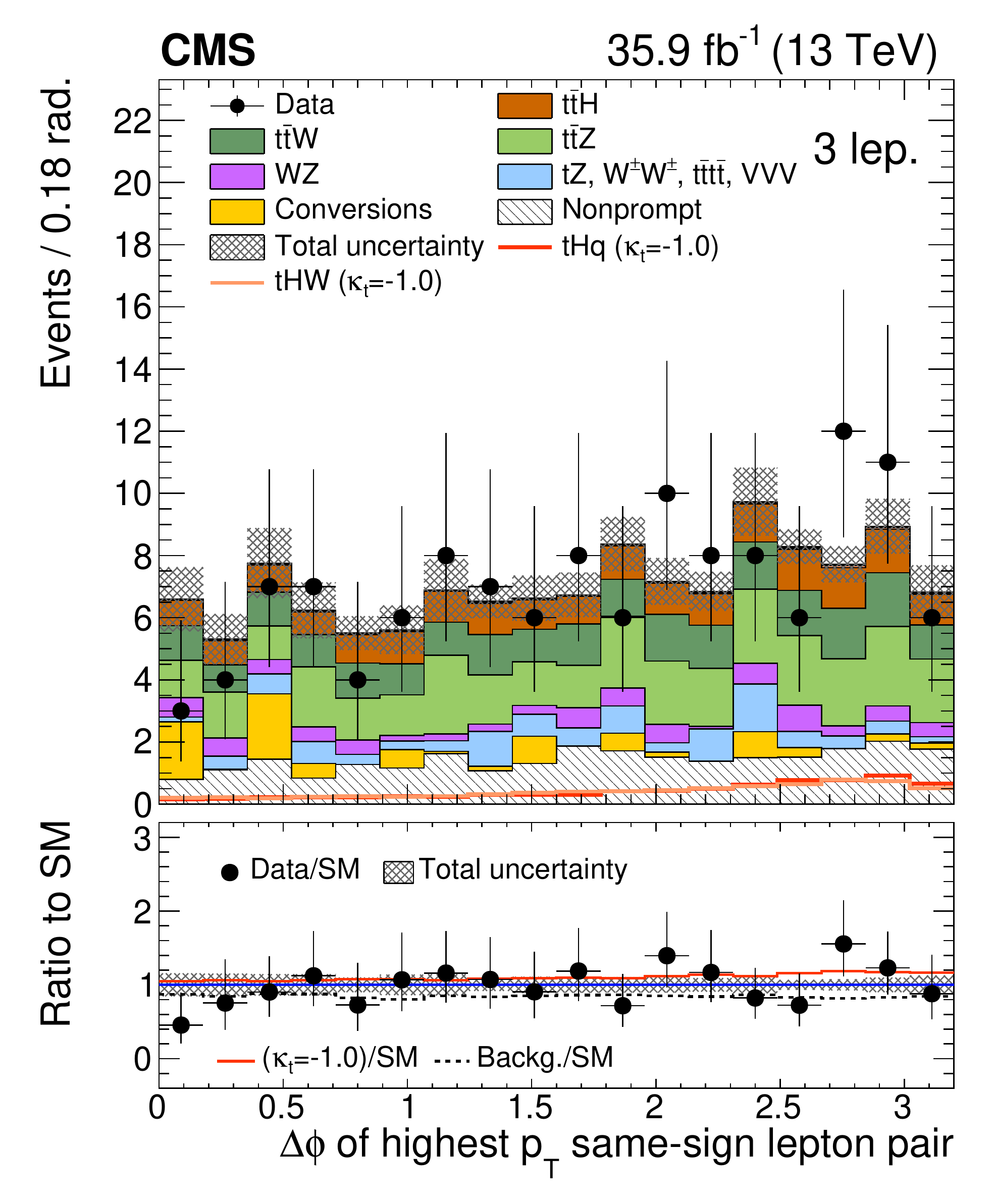}
    \includegraphics[width=0.32\linewidth]{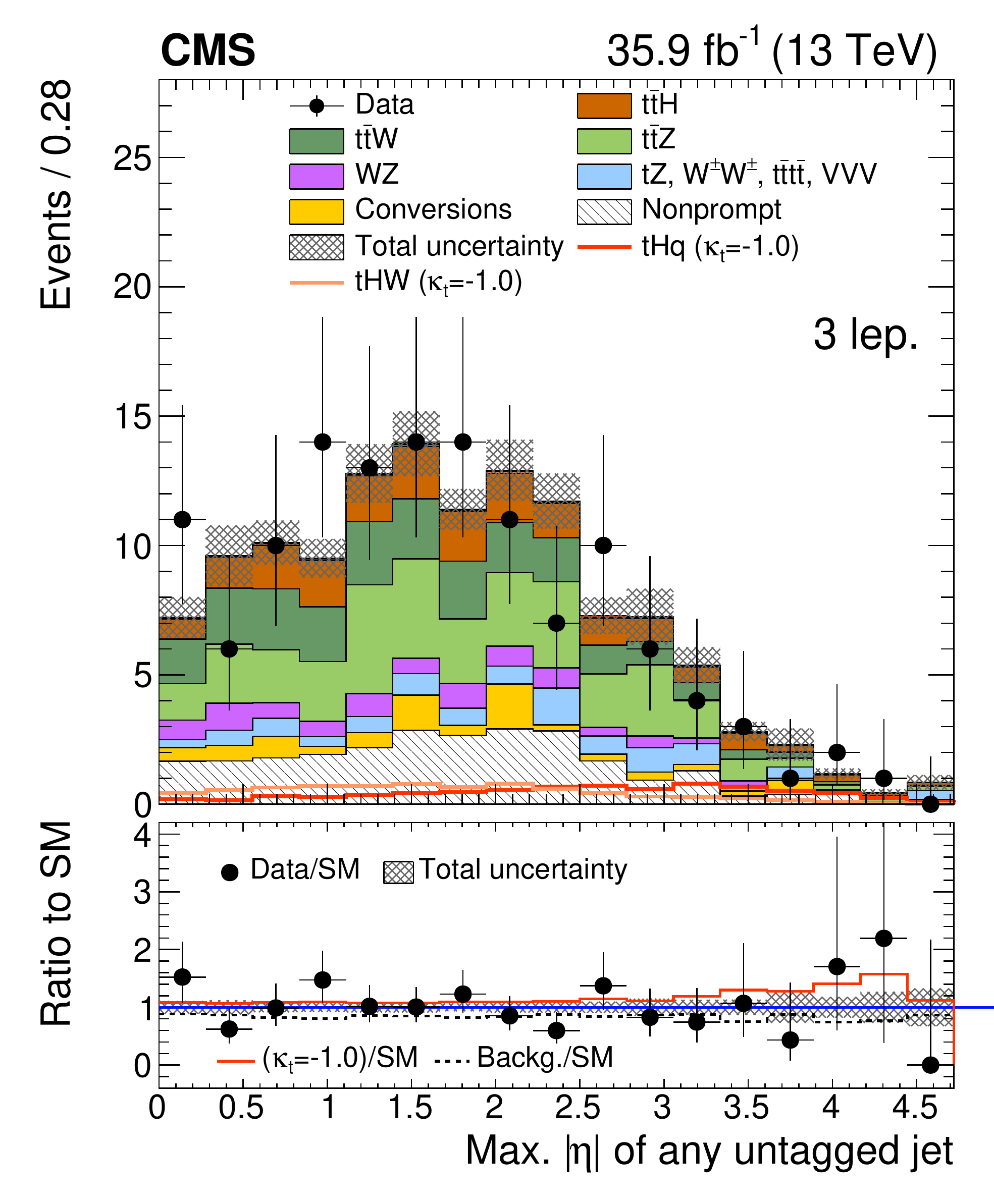}
    \includegraphics[width=0.32\linewidth]{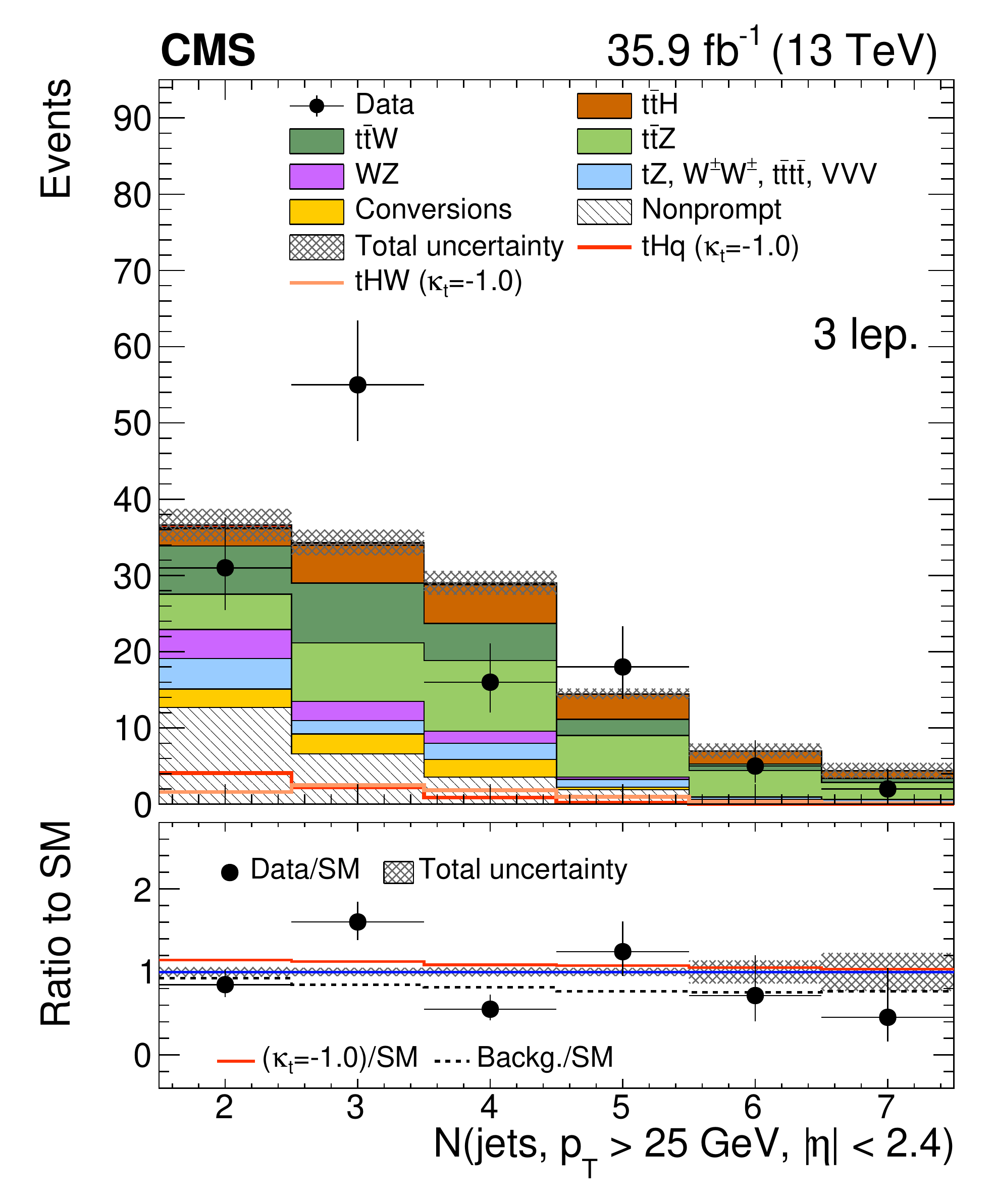} \\
  \caption{Distributions of discriminating observables for the three lepton channel, normalized to 35.9\fbinv, before fitting the signal discriminant to the data. The grey band represents the unconstrained (pre-fit) statistical and systematic uncertainties. In the panel below each distribution, the ratio of the observed and predicted event yields is shown. The shape of the two \tH\ signals for $\kappat=-1.0$ is shown, normalized to their respective cross sections for $\kappat=-1.0, \kappaV=1.0$.\label{fig:2lss_inputs_3l}}
\end{figure*}

A selection of the main discriminating input observables is shown in Figs.~\ref{fig:2lss_inputs_mm}--\ref{fig:2lss_inputs_3l}, comparing the data and the estimated distribution of signal and background processes.


\begin{figure*}[htb]
  \begin{center}
    \includegraphics[width=0.32\textwidth]{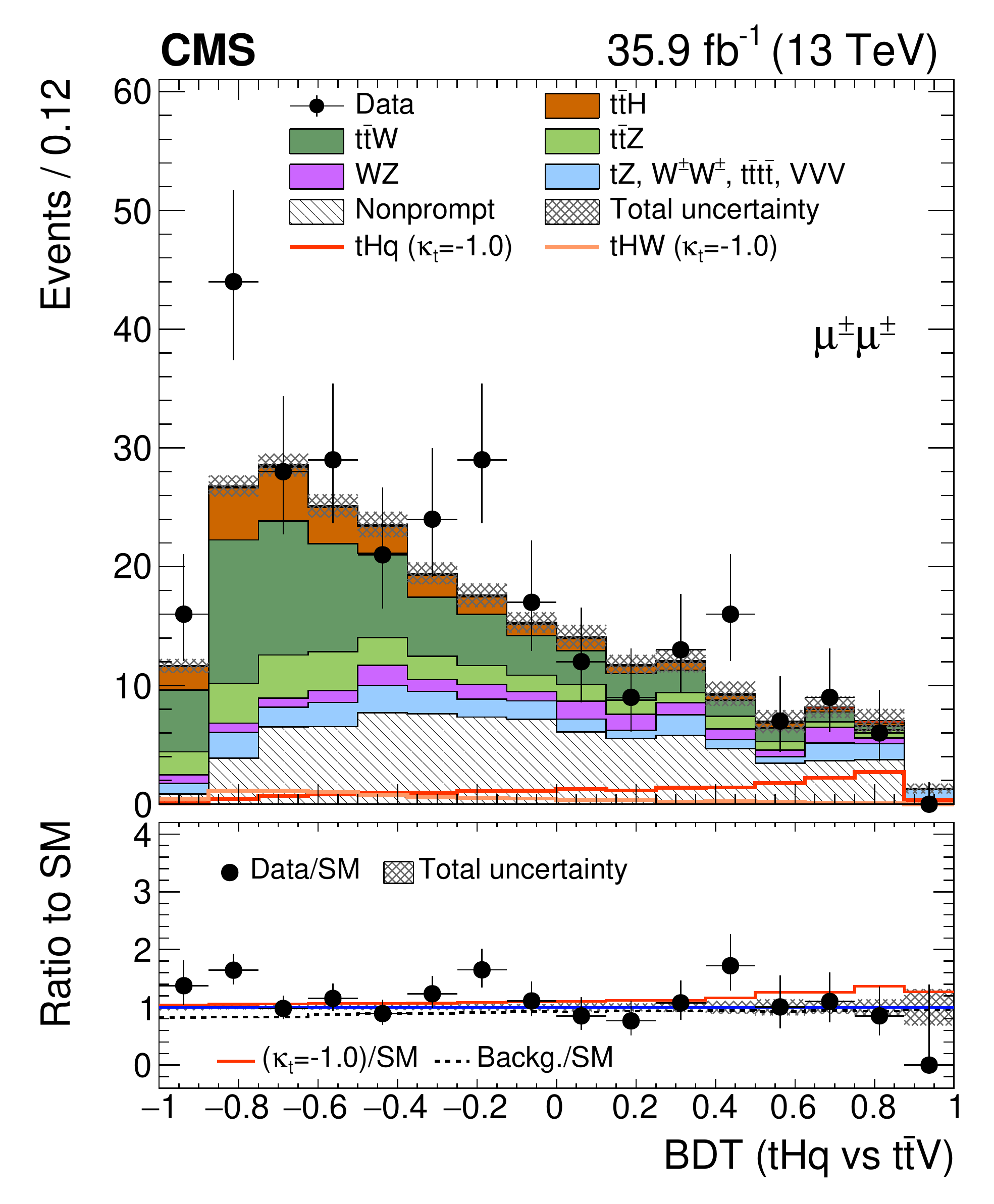}
    \includegraphics[width=0.32\textwidth]{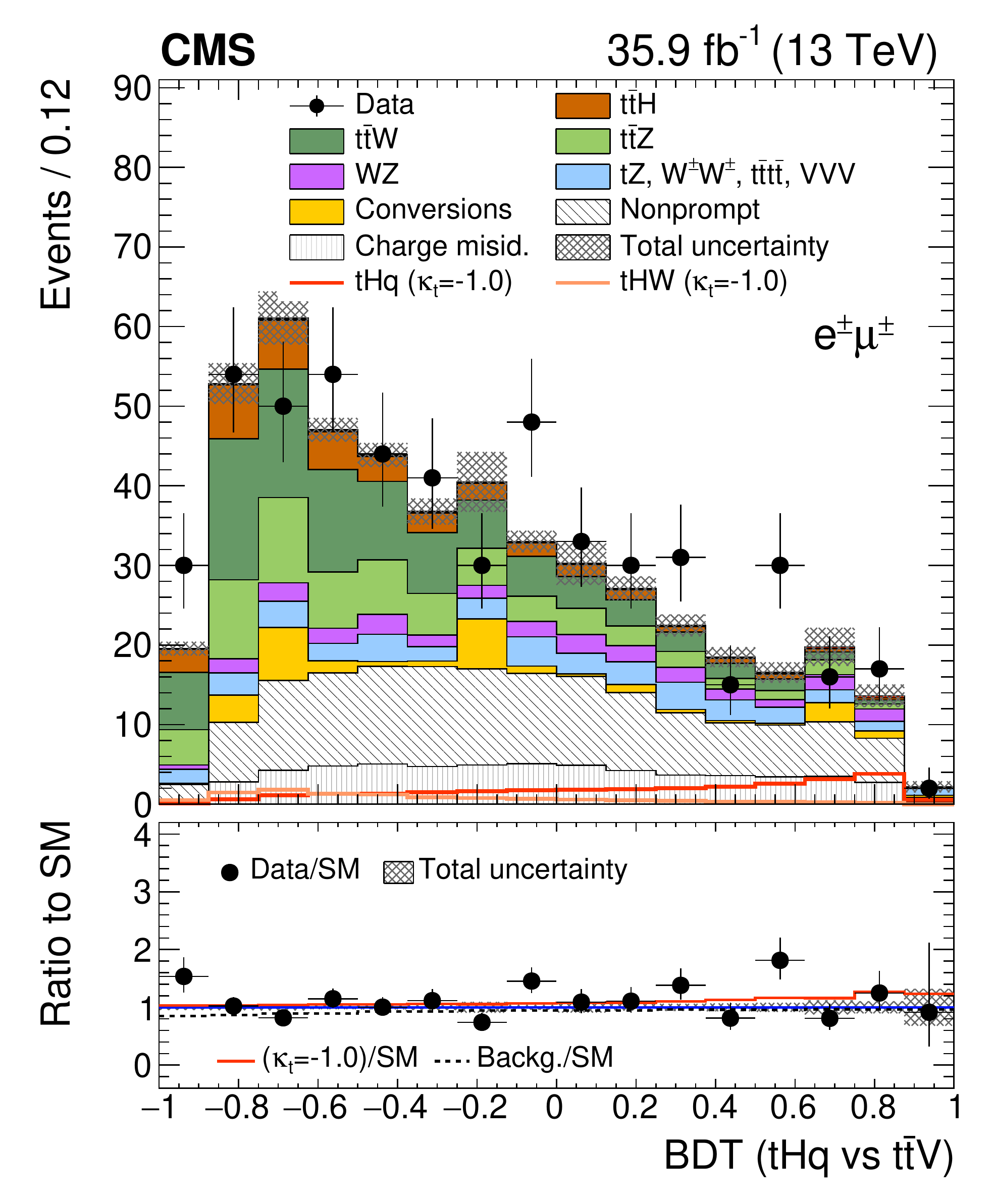}
    \includegraphics[width=0.32\textwidth]{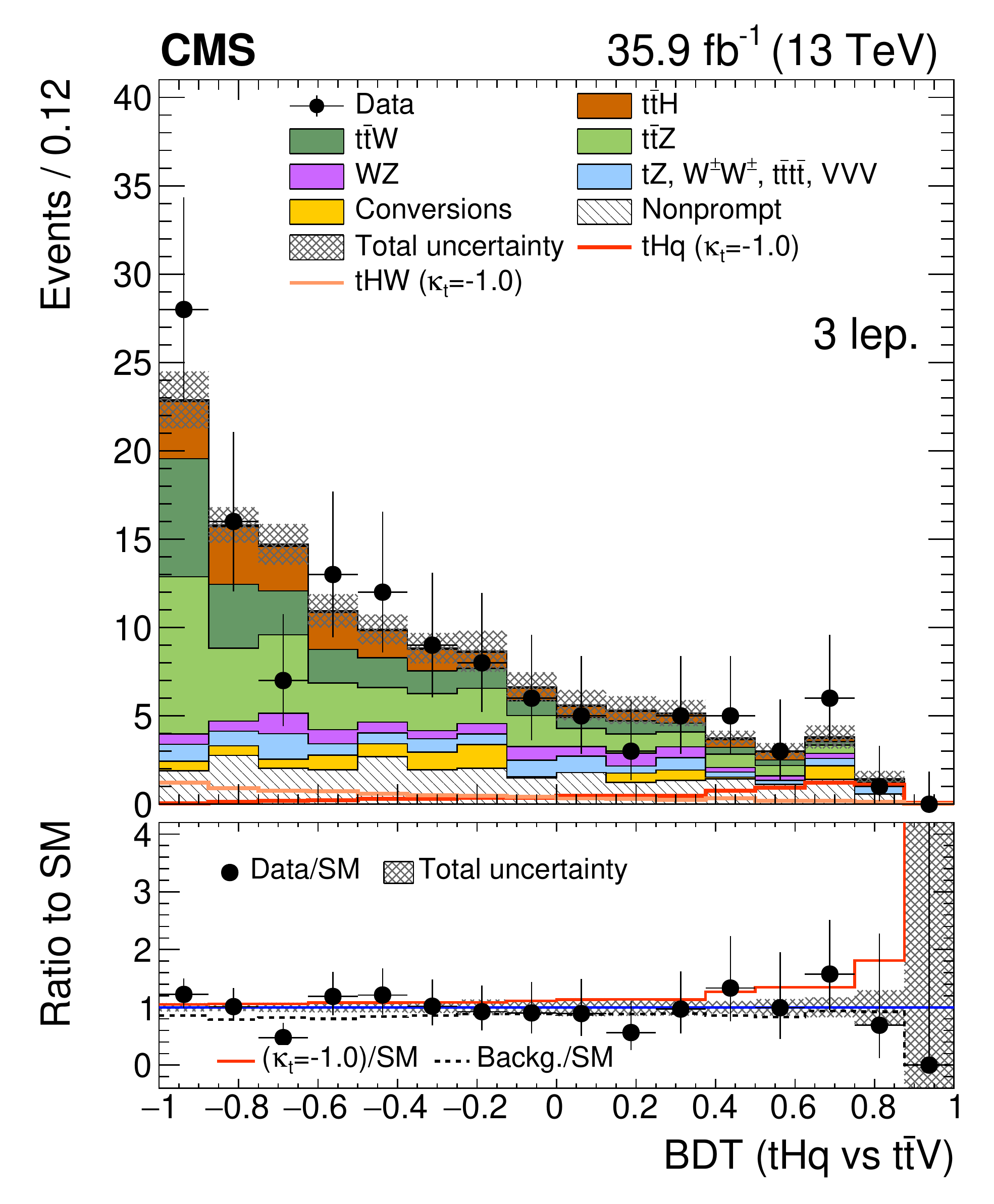}  \\
    \includegraphics[width=0.32\textwidth]{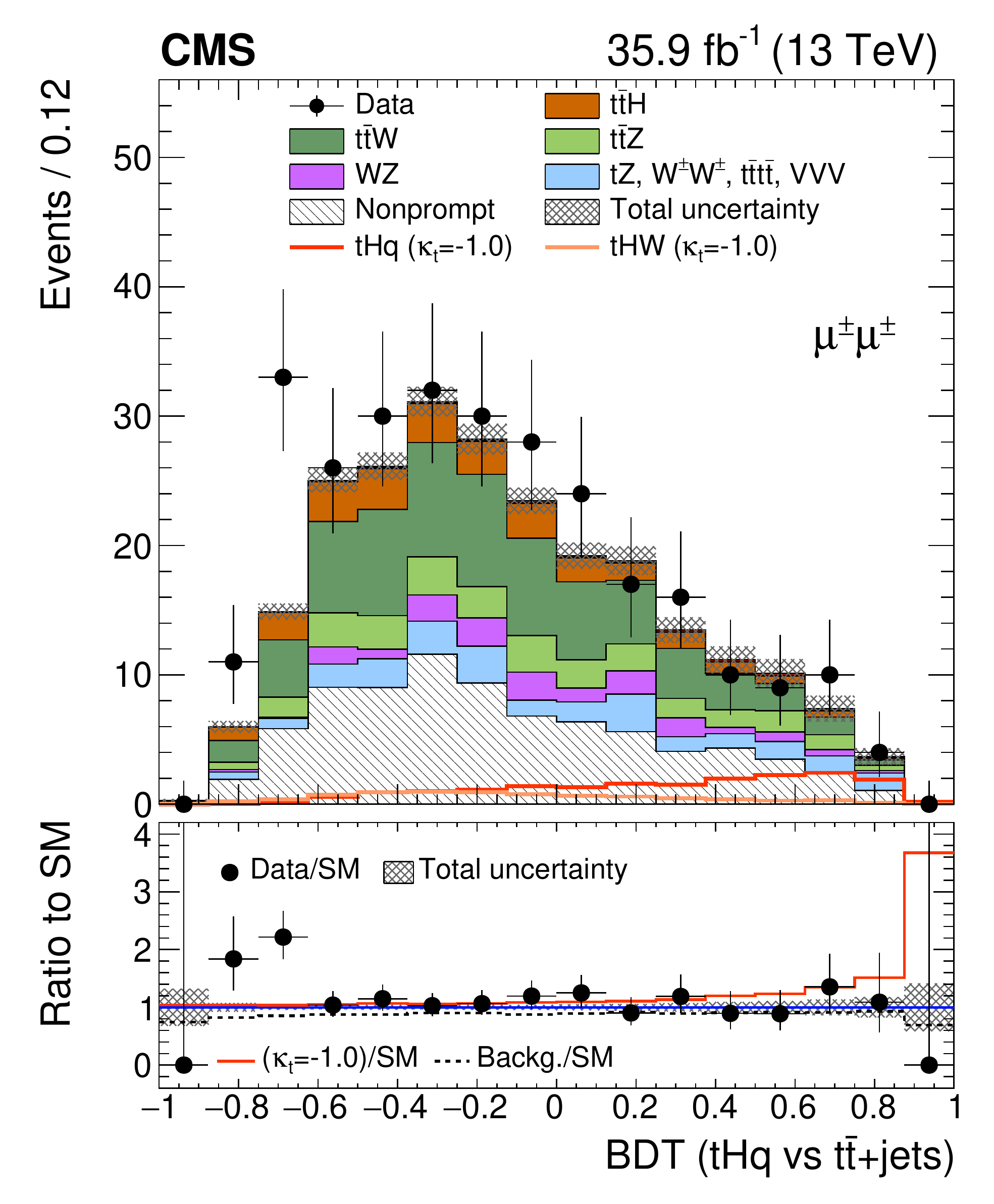}
    \includegraphics[width=0.32\textwidth]{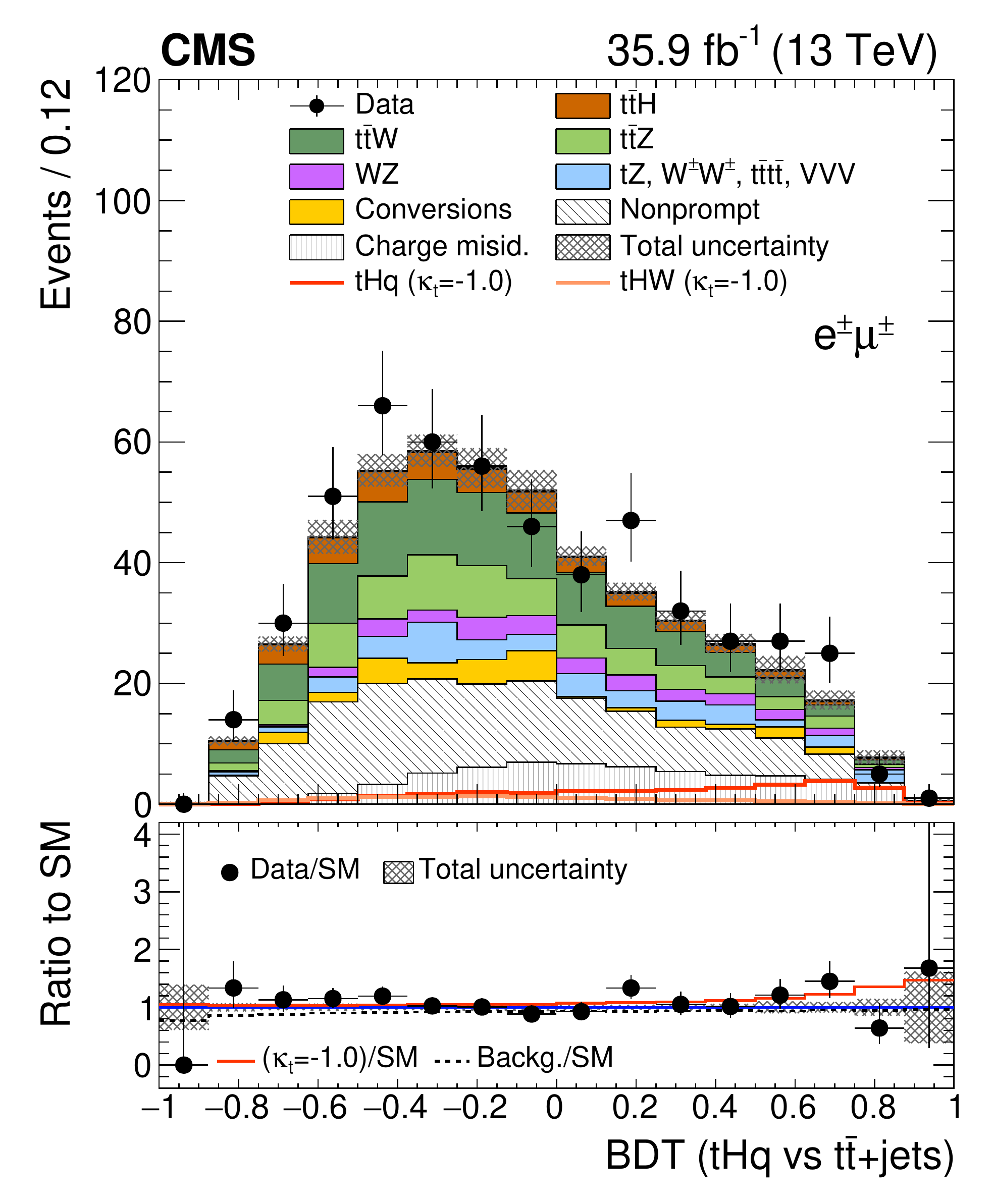}
    \includegraphics[width=0.32\textwidth]{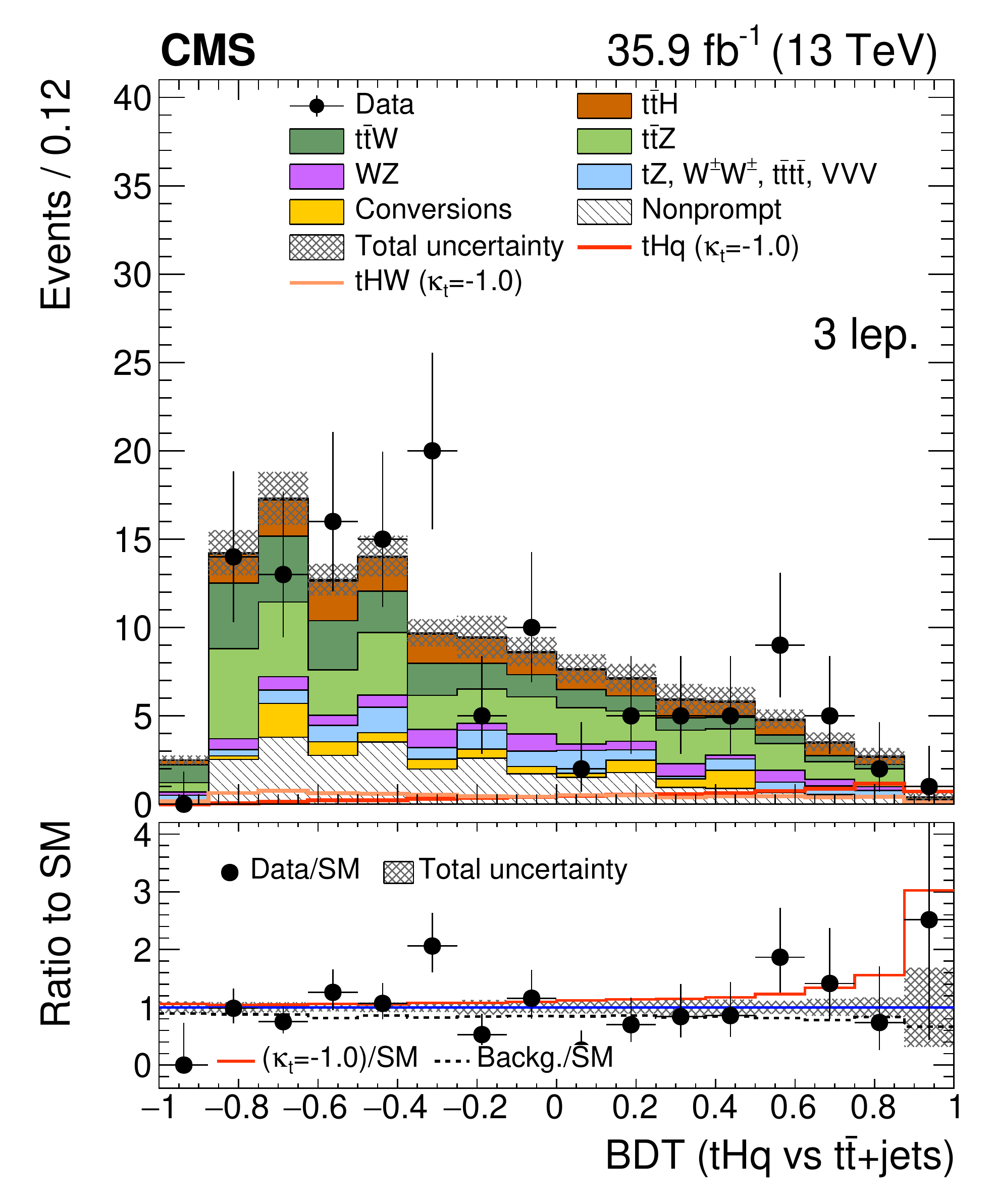}
  \end{center}
  \caption{Pre-fit classifier outputs, for the \mupmup\ channel (left), \epmup\ channel (center), and three-lepton channel (right), for training against \ttV\ (top row) and against \ttjets\ (bottom row).
  In the box below each distribution, the ratio of the observed and predicted event yields is shown.
  The shape of the two \tH\ signals for $\kappat=-1.0$ is shown, normalized to their respective cross sections for $\kappat=-1.0, \kappaV=1.0$.
  The grey band represents the unconstrained (pre-fit) statistical and systematic uncertainties.\label{fig:mlpostfitshapes}}
\end{figure*}

\begin{figure*}[htb]
  \begin{center}
    \includegraphics[width=0.32\textwidth]{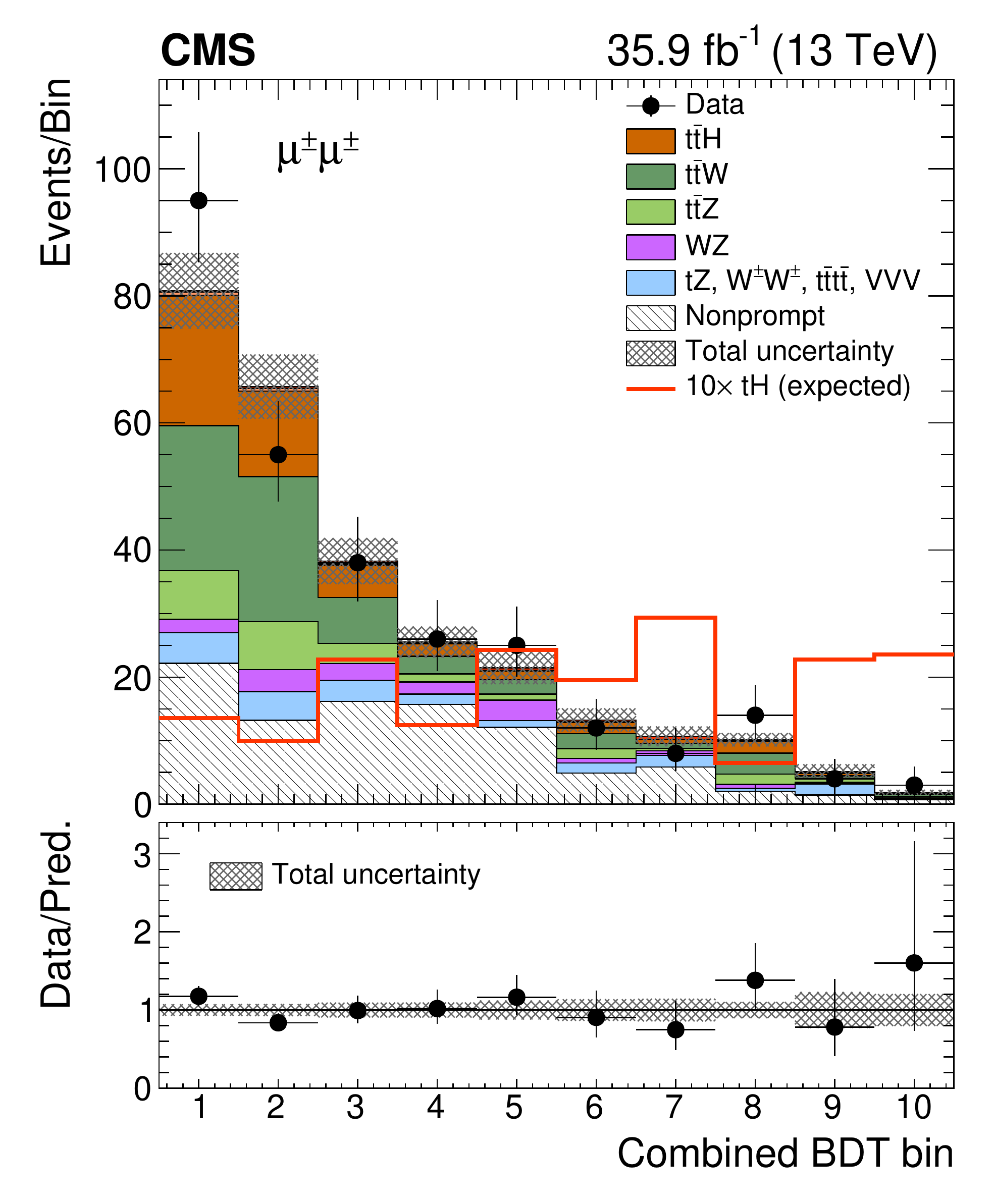}
    \includegraphics[width=0.32\textwidth]{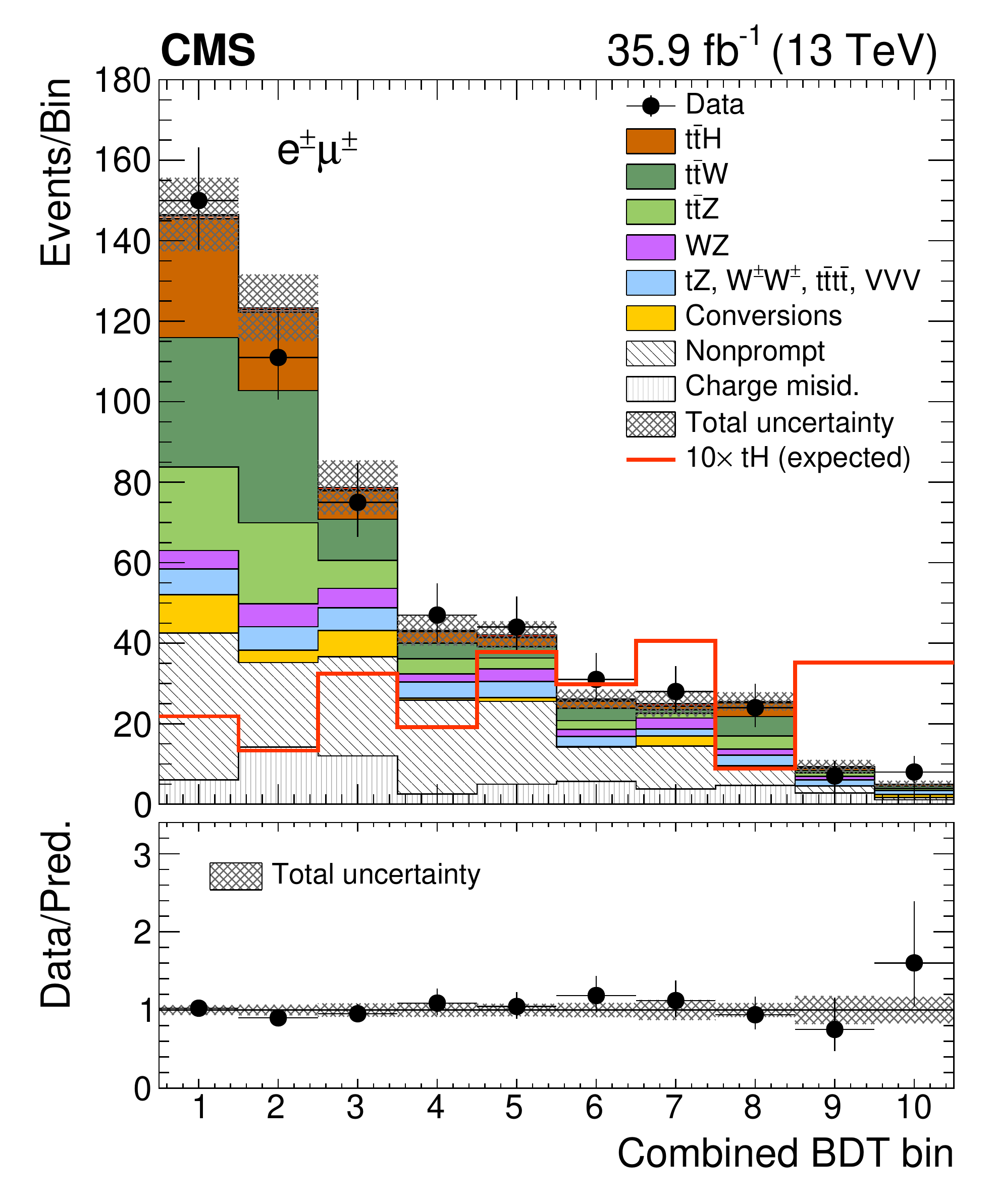}
    \includegraphics[width=0.32\textwidth]{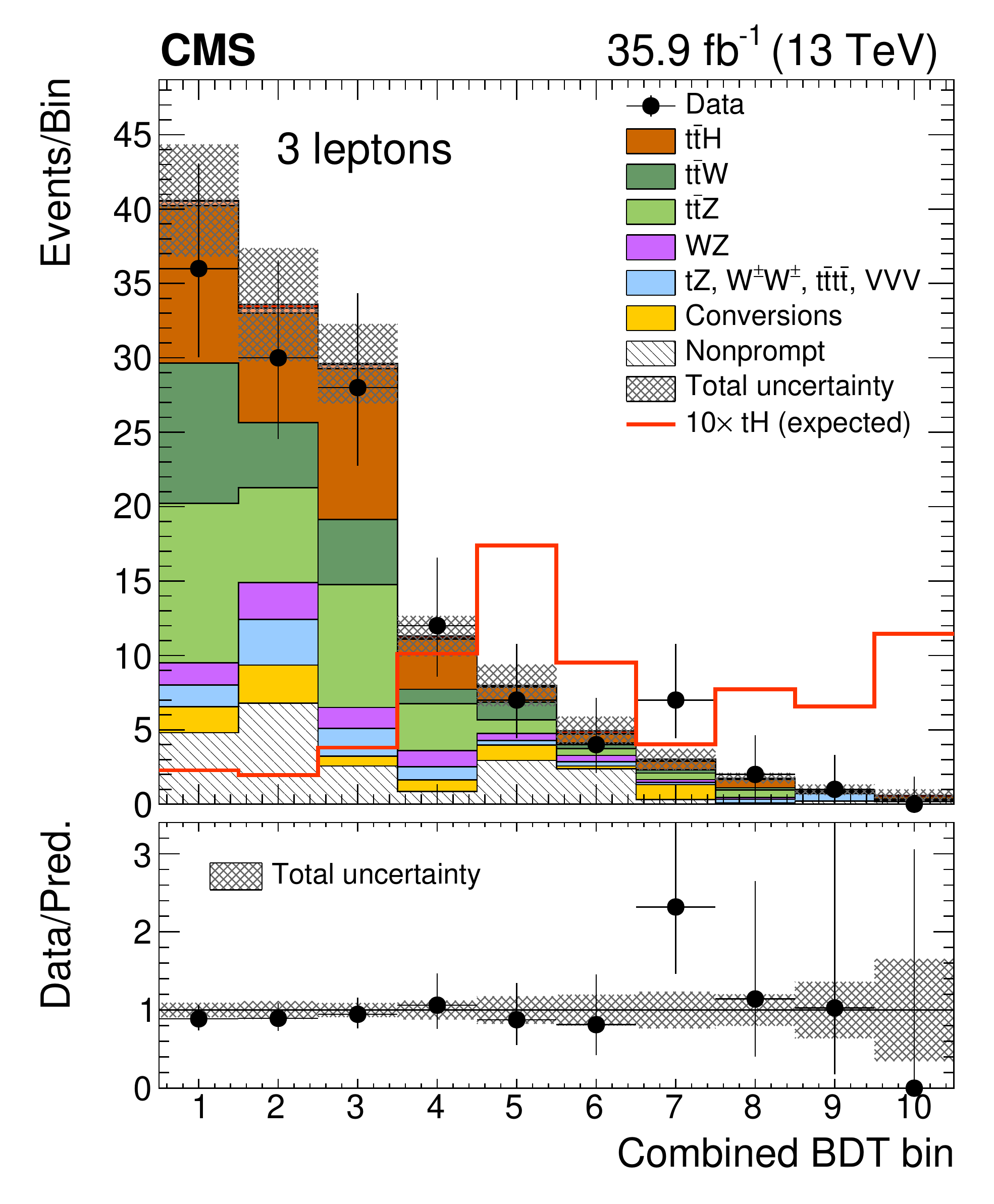}
  \end{center}
  \caption{Post-fit categorized classifier outputs as used in the maximum likelihood fit for the \mupmup\ channel (left), \epmup\ channel (center), and three-lepton channel (right), for 35.9\fbinv.
  In the box below each distribution, the ratio of the observed and predicted event yields is shown.
  The shape of the \tH\ signal is indicated with 10 times the SM.
  \label{fig:mlfinalbins}}
\end{figure*}

The six classifier output distributions, trained against \ttV\ and \ttjets\ processes for each of the three channels, are shown in Fig.~\ref{fig:mlpostfitshapes}, before a fit to the data.
The events are then sorted into ten categories depending on the output of the two BDT classifiers according to an optimized binning strategy, resulting in a one-dimensional histogram with ten bins.
Figure~\ref{fig:mlfinalbins} shows the post-fit categorized classifier output distributions for each of the three channels, after the combined maximum likelihood fit to extract the limits, as described in Section~\ref{sec:results}.

\subsection{Systematic uncertainties}\label{ssec:multilepsystematics}
The yield of signal and background events after the selection, as well as the shape of the classifier output distributions for signal and background processes, have systematic uncertainties from a variety of sources, both experimental and theoretical.
Experimental uncertainties relate either to the reconstruction of physics objects or to imprecisions in estimating the background contributions.
Uncertainties in the efficiency of reconstructing and selecting physics objects affect all yields and kinematic shapes taken from MC simulation, for both signal and background.
Background contributions estimated from the data are not affected by these.

Uncertainties from unknown higher-order contributions to \tHq\ and \tHW\ production are estimated from a change in the factorization and renormalization scales of double and half the initial value, evaluated separately for each point of \kappat.
The \ttH\ component has an uncertainty of between 5.8--9.3\% from scale variations and an additional $3.6\%$ from the knowledge of PDFs and the strong coupling constant $\alpS$~\cite{deFlorian:2016spz}.
Uncertainties related to the choice of the PDF set and its scale are estimated to be $3.7\%$ for \tHq\ and $4.0\%$ for \tHW.
The effect of missing higher-order corrections to the kinematic shape of the classifier outputs is taken into account for the \tH, \ttH, and \ttV\ components by independent changes of the renormalization and factorization scales of double and half the nominal value.

The cross sections of \ttZ\ and \ttW\ production are known with uncertainties of $+9.6\%/\!-11.2\%$ and $+12.9\%/\!-11.5\%$, respectively, from missing higher-order corrections to the perturbative expansion.
The corresponding values due to uncertainties in the PDFs and $\alpS$ are $3.4$ and $4.0\%$, respectively~\cite{deFlorian:2016spz}.

The efficiency for events passing the combination of trigger requirements is measured separately for events with two or more leptons, and has an uncertainty in the range of 1--3\%.
Efficiencies for the reconstruction and selection of muons and electrons are measured as a function of their \pt, using a tag-and-probe method with uncertainties of 2--4\%~\cite{EWK-10-002}.
The energy scale of jets is determined using event balancing techniques and carries uncertainties of a few percent, depending on \pt\ and $\eta$ of the jets~\cite{Khachatryan:2016kdb}.
Their impact on the kinematic distributions used in the signal extraction are estimated by varying the scales within their respective uncertainty and propagating the effects to the final result, recalculating all kinematic quantities and reapplying the event selection criteria.
The \cPqb\ tagging efficiencies are measured in heavy-flavor enriched multijet events and in \ttbar\ events, with \pt- and $\eta$-dependent uncertainties of a few percent~\cite{Sirunyan:2017ezt}.

The uncertainty in the integrated luminosity is 2.5\%~\cite{LUM-17-001} and affects the normalization of all processes modeled in simulation.

The estimate of events containing nonprompt leptons is subject to uncertainties in the determination of the tight-to-loose ratio on one hand and to the inherent bias in the selection of the control region dominated by nonprompt leptons, as tested in simulated events, on the other hand.
The measurement of the lepton tight-to-loose rate has statistical as well as systematic uncertainties from the removal of residual prompt leptons in the control region, amounting to a total uncertainty of  10--40\%, depending on the flavor of the leptons and their \pt\ and $\eta$.
The validity of the method itself is tested in simulated events and contributes a small additional uncertainty both to the normalization and shape of the classifier distributions for such events.

The estimate of backgrounds from electron charge misidentification in the \epmup\ channel carries an uncertainty of about 30\% from the measurement of the misidentification probability.
It is composed of a dominant statistical component from the limited event yields, and one related to the residual disagreement observed when testing the prediction in a control region.

The estimate of backgrounds from \WZ\ production is normalized in a control region with three leptons and carries uncertainties due to its limited statistics (10\%), the residual non-\WZ\ backgrounds (20\%), the \cPqb\ tagging rate ($10\%$), and the theoretical uncertainties related to the flavor composition of jets produced in association with the boson pair (up to 10\%).
In the dilepton channels, this uncertainty is increased to 50\% to account for the differences with respect to the control region.

Additional smaller backgrounds which have not yet been observed at the LHC (VVV, same-sign \PW\ boson production, \tZq, \tZW, \tttt) are assigned a normalization uncertainty of 50\%.

Of these sources of systematic uncertainties, the ones with largest impact on the final result are found to be those related to the normalization of the nonprompt backgrounds, the scale variations for the \ttV\ and \ttH\ processes, and the lepton selection efficiencies.

\section{Single-lepton + \texorpdfstring{\bbbar}{bbbar} channels}\label{sec:bbchannels}
Events from a \tH\ signal where the Higgs boson decays to a bottom quark-antiquark pair (\Hbb) produce final states with at least three central \cPqb\ jets and a hard lepton from the top quark decay chain used for triggering.
Selecting such events leads to challenging backgrounds from \ttbar\ production with additional heavy-flavor quarks, which can be produced in gluon splittings from initial- or final-state radiation.
The analysis described in this section uses two selections aimed at identifying signal events, with either three or four \cPqb-tagged jets, and a separate sample with opposite-sign dileptons, dominated by \ttjets\ events, to control $\ttbar+\text{heavy-flavor}$ (\tthf) events in a simultaneous fit.
A multivariate classification algorithm is trained to discriminate different \ttjets\ background components in the control region.
Additional multivariate algorithms are used to optimize the jet-parton assignment used to reconstruct kinematic properties of signal and background events which, in turn, are used to distinguish these components.

\subsection{Selection}\label{ssec:bbselection}
Selected events in the single-lepton + \bbbar\ signal channels must pass a single-lepton trigger.
Each event is required to contain exactly one muon or electron.
Muon (electron) candidates are required to satisfy $\pt > 27$ $(35)\GeV$ and $\abs{\eta} < 2.4$ $(2.1)$, motivated by the trigger selection, and to be isolated and fulfill strict quality requirements.
Events with additional leptons that have $\pt > 15\GeV$ and pass less strict quality requirements are rejected.
At the analysis level, the selection criteria target the \Hbb\ and $\cPqt\to\PW\cPqb\to\ell\nu\cPqb$ decay channels.
With these decays, the final state of the \tHq\ process consists of one \PW\ boson, three \cPqb\ quarks, and the light-flavor quark recoiling against the top quark-Higgs boson system.
In addition, a fourth \cPqb\ quark is expected because of the initial gluon splitting, but often falls outside the detector acceptance.
The main signal region is therefore required to have either three or four \cPqb-tagged jets and at least one additional untagged jet, both defined using the medium working point.
Central jets with $\abs{\eta} < 2.4$ are required to have $\pt > 30\GeV$, while jets in the forward region ($2.4 \leq \abs{\eta} \leq 4.7$) are required to have $\pt > 40\GeV$.

The neutrino is accounted for by requiring a minimal amount of missing transverse momentum of $\ptmiss>35\GeV$ in the muon channel and $\ptmiss>45\GeV$ in the electron channel.
This renders the background from QCD multijet events negligible.

In addition to the signal regions, a control region is defined to constrain the main background contribution from top quark pair production.
Events selected for this control region must pass a dilepton trigger.
Each event is required to contain exactly two oppositely charged leptons, where their flavor can be any combination of muons or electrons.
Two jets in each event must be \cPqb\ tagged.
Furthermore, at least one additional jet must pass the loose \cPqb\ tagging requirement.
Similarly to the signal regions, each event is further required to have a minimum amount of missing transverse momentum.
All selection criteria are summarized in Table~\ref{tab:bb_cuts}.

\begin{table}[!h]
  \topcaption{Summary of event selection for the single-lepton + \bbbar\ channels.\label{tab:bb_cuts}}
  \centering
  \begin{scotch}{p{8cm}}
    \quad Signal region \\
    One muon (electron) with $\pt>27 (35)\GeV$ \\
    No additional loose leptons with $\pt>15\GeV$ \\
    Three or four medium \cPqb-tagged jets     \\
    $\pt>30\GeV$ and $\abs{\eta}<2.4$          \\
    One or more untagged jets                  \\
    $\pt>30\GeV$ for $\abs{\eta}<2.4$ or       \\
    $\pt>40\GeV$ for $\abs{\eta}\ge2.4$        \\
    $\ptmiss>35 (45)\GeV$ for muons (electrons) \\
    [\cmsTabSkip]
    \quad Control region \\
    Two leptons: $\pt>20/20\GeV$ ($\Pgm^\pm\Pgm^\mp$) \\
    or $\pt>20/15\GeV$ ($\Pepm\Pe^\mp/\Pgm^\pm\Pe^\mp$)\\
    No additional loose leptons        \\
    with $\pt>20/15\GeV$ ($\Pgm^\pm/\Pepm$) \\
    Two medium \cPqb-tagged jets       \\
    $\pt>30\GeV$ and $\abs{\eta}<2.4$  \\
    One or more additional loose \cPqb-tagged jets \\
    $\pt>30\GeV$ and $\abs{\eta}<2.4$ \\
    $\ptmiss>40\GeV$\\
    \end{scotch}
\end{table}

\subsection{Backgrounds}\label{ssec:bbbackgrounds}
The main background contribution in the single-lepton + \bbbar\ channels arises from SM processes with multiple \cPqb\ quarks.
The modeling and estimation of all background processes are done using samples of simulated events.

In particular, the dominant background process is top quark pair production because of the similar final state and, comparatively, a large cross section.
As the modeling of the additional heavy-flavor partons in \ttbar\ events is theoretically difficult, the sample of simulated \ttbar events is further divided into different subcategories, defined by the flavor of possible additionally radiated quarks and taking into account a possible merging of \cPqb\ hadrons into single jets.
The control region is specifically designed to separate the \tthf\ and $\ttbar+\text{light-flavor}$ (\ttlf) components with a multivariate approach.
The different categories are listed in Table~\ref{tab:ttjetcats}.

\begin{table}[htb]
  \topcaption{Subcategories of \ttjets\ backgrounds used in the analysis.\label{tab:ttjetcats}}
  \centering
  \begin{scotch}{ll}
  \ttbb   & Two additional jets arising from \cPqb\ hadrons \\
  \tttwob & One additional jet arising from two merged \\
          & \cPqb\ hadrons \\
  \ttb    & One additional jet arising from one \cPqb\ hadron \\
  \ttcc   & The three former categories combined for \cPqc\ hadrons \\
          & instead of \cPqb\ hadrons \\
  \ttlf   & All events that do not meet the criteria of the other \\
          & four categories \\
  \end{scotch}
\end{table}

Other backgrounds contributing to the signal region are single top quark production and top quark pair production in association with electroweak bosons, namely \ttW\ and \ttZ.
An irreducible background for the \tHq\ processes comes from \tZq\ production with $\PZ\to\bbbar$.
Background contributions also arise from \Zjets production, especially in the dilepton control region.

The expected and observed event yields for the signal and control regions are listed in Table~\ref{tab:bb_yields}.

\begin{table*}[htb]
  \topcaption{Data yields and expected backgrounds after the event selection for the two signal regions and in the dilepton control region. Uncertainties include both systematic and statistical components.\label{tab:bb_yields}}
  \centering
  \begin{scotch}{lrrr}
      Process & 3 tags & 4 tags & Dilepton \\
    \hline
    \ttlf                         & $24100 \pm 5800$ & $320  \pm 180$ & $5300 \pm 1000$ \\
    \ttcc                         & $8500  \pm 4900$ & $340  \pm 260$ & $2100 \pm 1200$ \\
    \ttbb                         & $4100  \pm 2300$ & $780  \pm 430$ & $750  \pm 440 $ \\
    \ttb                          & $4000  \pm 2100$ & $180  \pm 110$ & $770  \pm 430 $ \\
    \tttwob                       & $2300  \pm 1200$ & $138  \pm 88 $ & $400  \pm 230 $ \\
    Single top                    & $1980  \pm 350 $ & $78   \pm 26 $ & $285  \pm 37  $ \\
    \ttZ                          & $202   \pm  30 $ & $32.0 \pm 6.6$ & $54.8 \pm 7.3 $ \\
    \ttW                          & $90    \pm 23  $ & $4.2  \pm 2.8$ & $31.4 \pm 5.9 $ \\
    \tZq                          & $28.3  \pm 5.7 $ & $2.9  \pm 2.3$ & \NA\            \\
    \Zjets\                       & \NA\             & \NA\           & $69   \pm 32$   \\
    [\cmsTabSkip]
    Total background              & $45300 \pm 8300$ & $1880 \pm 550$ & $9700 \pm 1700$ \\
    [\cmsTabSkip]
    \ttH                          & $268   \pm  31$  & $62.0 \pm 9.9$ & $48.9 \pm 5.9 $ \\
    $\tHq$ (SM)                   & $11.1  \pm 3.3$  & $ 1.3 \pm 0.3$ & $0.31 \pm 0.08$ \\
    $\tHW$ (SM)                   & $ 7.6  \pm 1.1$  & $ 1.1 \pm 0.3$ & $1.4  \pm 0.2 $ \\
    [\cmsTabSkip]
    Total SM                      & $45700 \pm 8300$ & $1940 \pm 550$ & $9700 \pm 1700$ \\
    [\cmsTabSkip]
    $\tHq$ ($\kappaV=1=-\kappat$) & $160   \pm 38$   & $19.1 \pm 5.2$ & $ 3.9 \pm  1.0$ \\
    $\tHW$ ($\kappaV=1=-\kappat$) & $ 92   \pm 12$   & $13.7 \pm 2.3$ & $17.6 \pm  2.2$ \\
    [\cmsTabSkip]
    Data                          & 44311            & 2035           & 9065  \\
  \end{scotch}
\end{table*}

\subsection{Signal extraction}\label{ssec:bbsignalextrac}
As the assignment of final state quarks to reconstructed jets is non-trivial for the multijet environment of the 3 and 4 tag signal regions, the jet-to-quark assignment is achieved with dedicated jet assignment BDTs (JA-BDTs).
Each event is reconstructed under three different hypotheses: \tHq\ signal event, \tHW\ signal event, or \ttjets\ background event.
Each assignment hypothesis utilizes a separate BDT, which is trained with correct and wrong jet-to-quark assignments of the respective process.
When a JA-BDT is applied, all possible jet-to-quark assignments are evaluated and the one with the highest JA-BDT output value is chosen for the given hypothesis.
The matching efficiency for a complete \tHq\ event is 58 (45)\% in the 3 (4) tag signal region, for a complete \tHW\ event 38 (29)\% and for a complete \ttbar\ event 58 (31)\%.

\begin{figure}[h]
  \centering
  \includegraphics[width=0.50\textwidth]{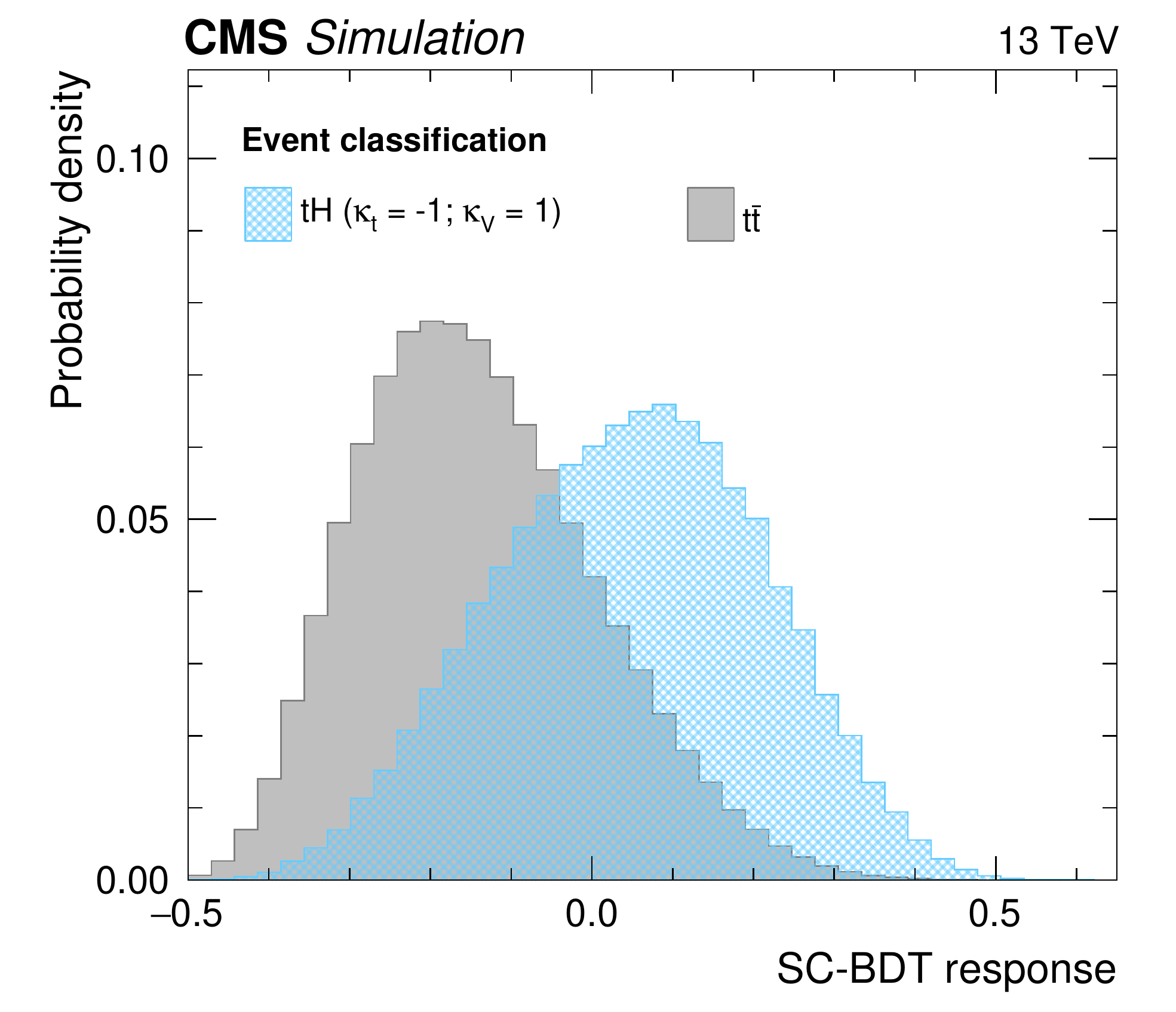}
  \caption{Output values of the SC-BDT.}
  \label{FigClassTHMVA}
\end{figure}

The different assignment hypotheses provide sensitive variables, which can be exploited in a further signal classification BDT (SC-BDT) to separate the \tHq\ and \tHW\ processes from the main background of the analysis, \ttbar\ events.
Global event variables that do not rely on any particular jet-to-quark assignment are used in addition to the assignment-dependent variables.
The input variables used for the SC-BDT are listed in Table~\ref{tab:class-vars} with the result of the training illustrated in Fig.~\ref{FigClassTHMVA}.
\begin{table*}[h!]
\renewcommand{\arraystretch}{1.3}
\center{
\topcaption[Classification variable description]{Description of variables used in the SC-BDT.
There are four types of variables: variables independent of any jet assignment, and variables based on objects obtained under the \ttbar, \tHq, or \tHW\ jet assignment. The natural logarithm transformation is used to smoothen and constrain broad distributions to a more narrow range.}
\begin{scotch}{p{0.35\textwidth}p{0.6\textwidth}}
 Variable  & Description \\
\hline
\quad Event variables                      & \\
$\ln{m_3}$                                 & Invariant mass of three hardest jets in the event\\
Aplanarity                                 & Aplanarity of the event~\cite{Barger:1993ww} \\
Fox--Wolfram \#1                           & First Fox--Wolfram moment~\cite{PhysRevLett.41.1581} of the event\\
$q(\ell)$                                  & Electric charge of the lepton \\
 [\cmsTabSkip]
\quad \ttbar\ jet assignment variables     & \\
$\ln{m(\thad)}$                            & Invariant mass of the reconstructed hadronically decaying top quark \\
CSV(\Whad jet 1)                           & Output of the \cPqb\ tagging discriminant for the first jet assigned to the hadronically decaying \PW\ boson \\
CSV(\Whad jet 2)                           & Output of the \cPqb\ tagging discriminant for the second jet assigned to the hadronically decaying \PW\ boson \\
$\Delta R$(\Whad jets)                     & $\Delta R$ between the two light jets assigned to the  hadronically decaying \PW\ boson \\
[\cmsTabSkip]
\quad \tHq\ jet assignment variables       & \\
$\ln{\pt(\PH)}$                            & Transverse momentum of the reconstructed Higgs boson candidate \\
$\abs{\eta(\text{light-flavor jet})}$      & Absolute pseudorapidity of light-flavor forward jet \\
$\ln{m(\PH)}$                              & Invariant mass of the reconstructed Higgs boson candidate\\
CSV(\PH jet 1)                             & Output of the \cPqb\ tagging discriminant for the first jet assigned to the Higgs boson candidate \\
CSV(\PH jet 2)                             & Output of the \cPqb\ tagging discriminant for the second jet assigned to the Higgs boson candidate \\
$\cos \theta(\text{b}_\text{t},\,\ell)$    & Cosine of the angle between the \cPqb-tagged jet from the top quark decay and the lepton\\
$\cos \theta^{*}$                          & Cosine of the angle between the light-flavor forward jet and the lepton in the top quark rest frame \\
$\abs{\eta(\text{t})$ - $\eta(\text{H}) }$ & Absolute pseudorapidity difference of reconstructed Higgs boson and top quark\\
$\ln{\pt(\text{light jet})}$               & Transverse momentum of the light-flavor forward jet \\
  [\cmsTabSkip]
\quad \tHW jet assignment variable         & \\
{JA-BDT response }                         & Best output of the \tHW\ JA-BDT\\
\end{scotch}
\label{tab:class-vars}
}
\end{table*}

\begin{figure}
  \centering
  \includegraphics[width=0.50\textwidth]{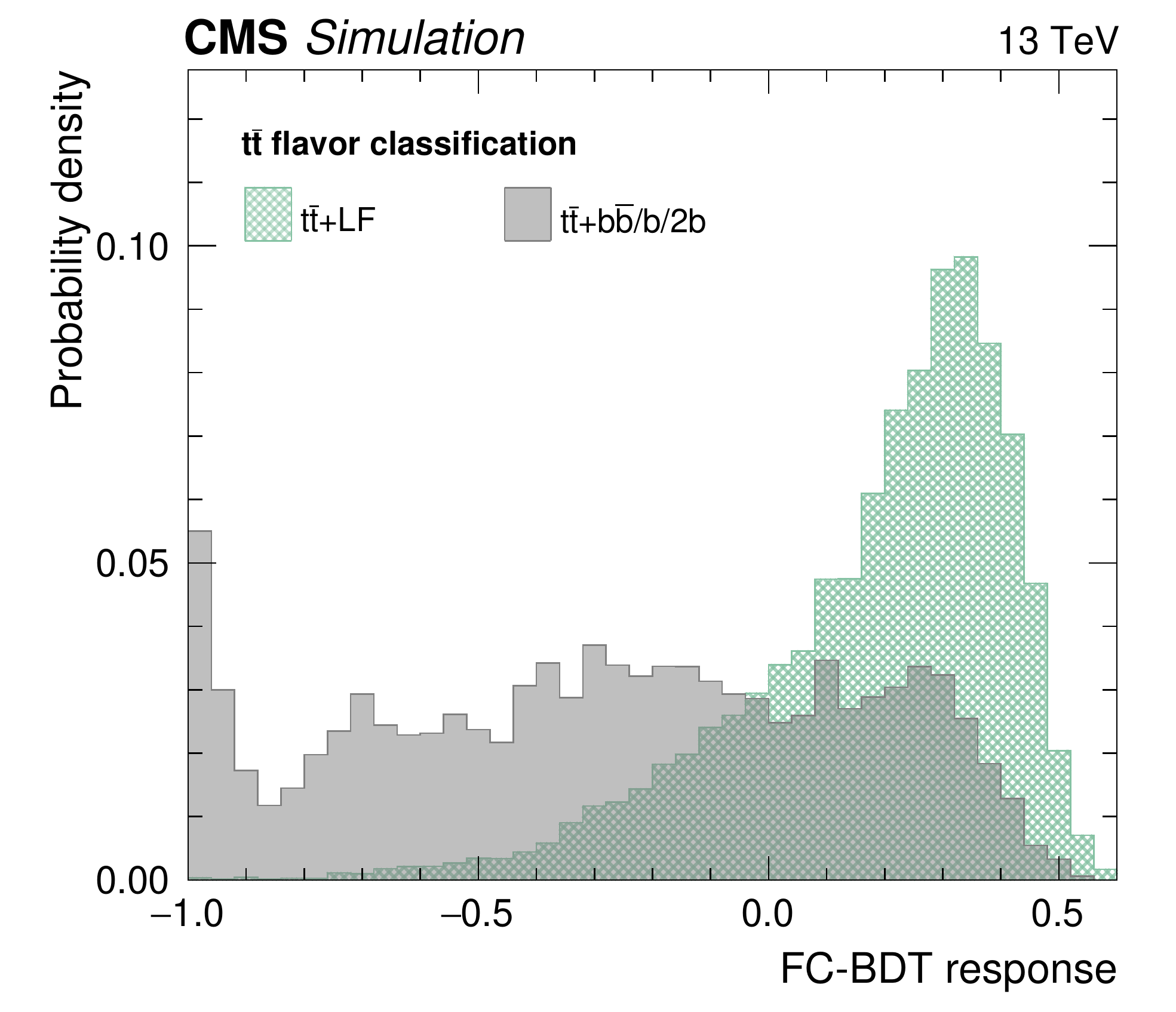}
  \caption{Response values of the FC-BDT. The background consists of \ttbb, $\ttbar+\mathrm{1\bar{b}}$, and $\ttbar+\mathrm{2\bar{b}}$ events.}
  \label{FigFlavorMVA}
\end{figure}

In addition, a dedicated flavor classification BDT (FC-BDT) is used in the dilepton region to constrain the contribution of different $\ttbar+\mathrm{X}$ components.
The training is performed with \ttlf\ as signal process and \ttbb\ as background process.
This FC-BDT exploits information of the number of jets per event and their response to \cPqb\ and \cPqc\ tagging algorithms.
The full list of input variables is provided in Table~\ref{TabFlavorVars} and the result of the training of the FC-BDT is shown in Fig.~\ref{FigFlavorMVA}.
\begin{table*}[tb]
\renewcommand{\arraystretch}{1.5}
\topcaption{Input variables used in the training of the FC-BDT. The variables are sorted by their importance in the training within each category. In total, eight variables are used for the training of the FC-BDT.}
\label{TabFlavorVars}
\small
\begin{scotch}{p{0.3\textwidth}p{0.65\textwidth}}
Variable                   & Description \\
\hline
{CSV(\cPqb jet 3)}         & Output of the \cPqb\ tagging discriminant for the \cPqb-tagged jet with the third-highest \cPqb\ tagging value in the event \\
{n$_{\text{jets}}$(tight)} & Number of jets in the event passing the tight working point of the \cPqb\ tagging algorithm \\
{CvsL(jet \pt 3)}          & Output of the charm \vs\ light-flavor tagging algorithm for the jet with the third-highest transverse momentum in the event \\
{CSV(\cPqb-tagged jet 2)}  & Output of the \cPqb\ tagging discriminant for the \cPqb-tagged jet with the second-highest \cPqb\ tagging value in the event \\
{CvsL(jet \pt 4})          & Output of the charm \vs\ light-flavor tagging algorithm for the jet with the fourth-highest transverse momentum in the event \\
{CvsB(jet \pt 3)}          & Output of the charm \vs\ bottom flavor tagging algorithm for the jet with the third-highest transverse momentum in the event \\
{CSV(\cPqb-tagged jet 4)}  & Output of the \cPqb\ tagging discriminant for the \cPqb-tagged jet with the fourth-highest \cPqb\ tagging value in the event \\
{n$_{\text{jets}}$(loose)} & Number of jets in the event passing the loose working point of the \cPqb\ tagging algorithm  \\
\end{scotch}
\end{table*}

\begin{figure*}[!htb]
\hspace{0.5cm}
\begin{center}
\includegraphics[width=0.48\textwidth]{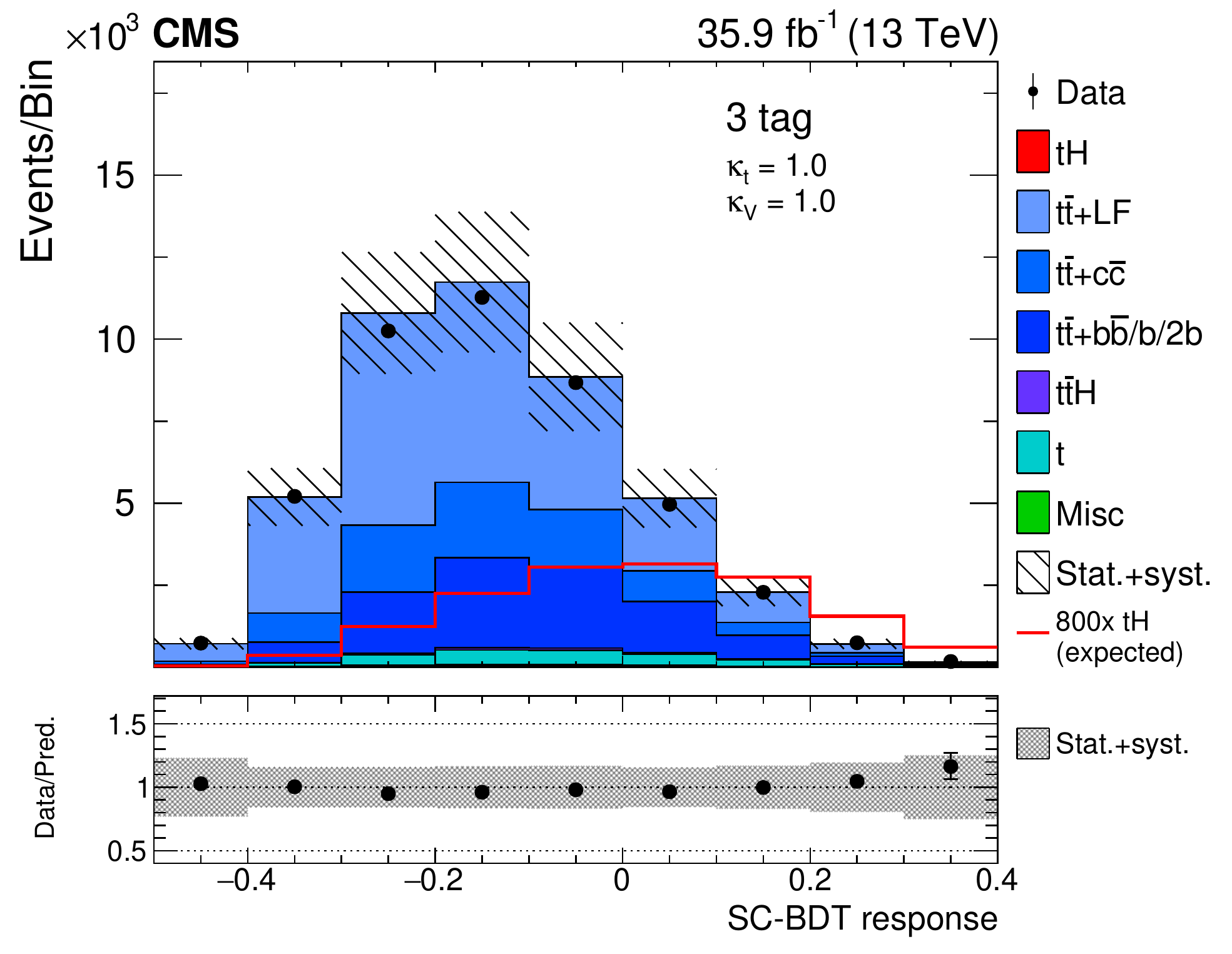}
\includegraphics[width=0.48\textwidth]{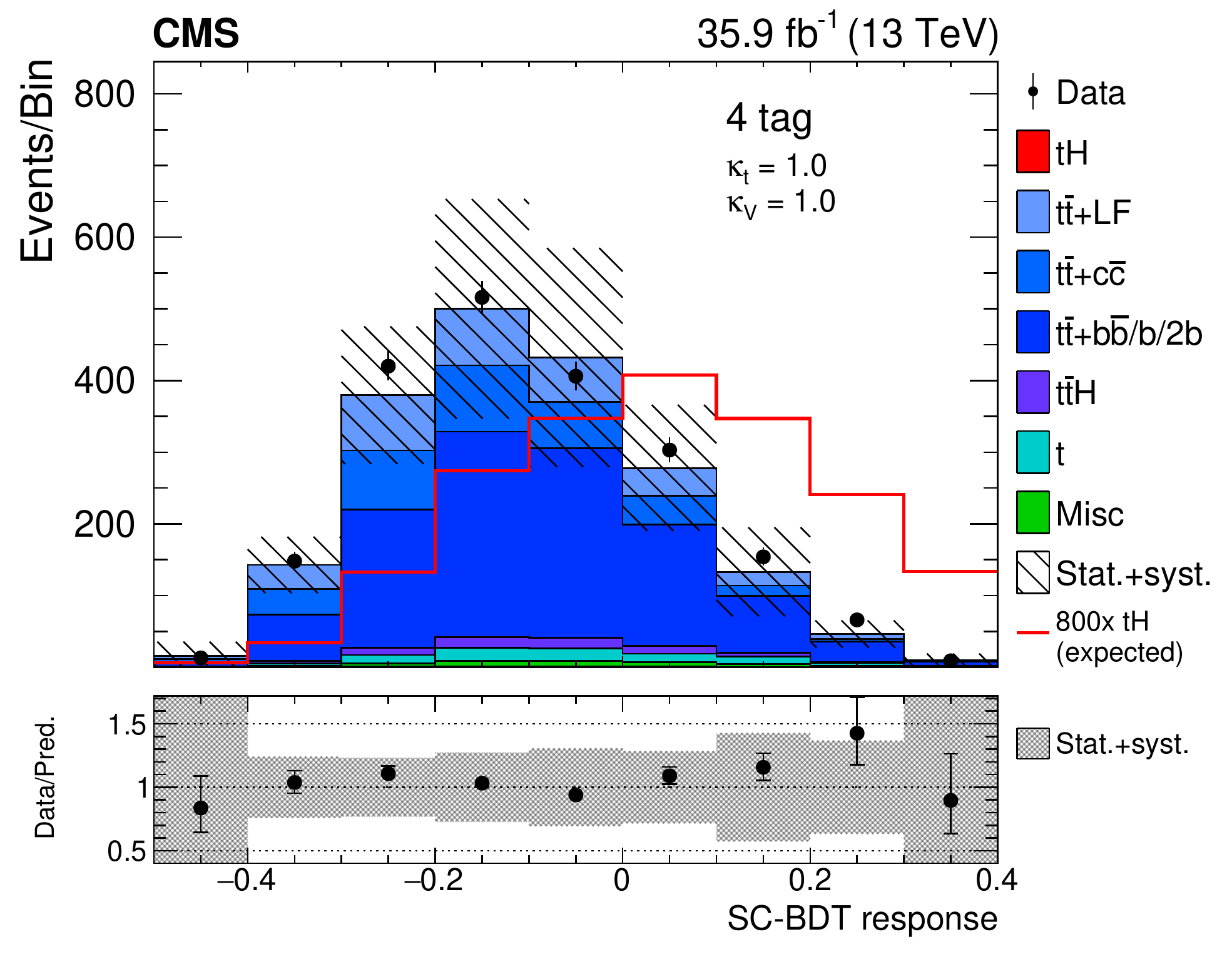}
\end{center}
\caption{Pre-fit classifier outputs of the signal classification BDT for the 3 tag channel (left) and the 4 tag channel (right), for 35.9\fbinv. In the box below each distribution, the ratio of the observed and predicted event yields is shown. The shape of the \tH\ signal is indicated with 800 times the SM.
\label{fig:bb_prefit}}
\end{figure*}

\begin{figure*}[!htb]
\begin{center}
\includegraphics[width=0.48\textwidth]{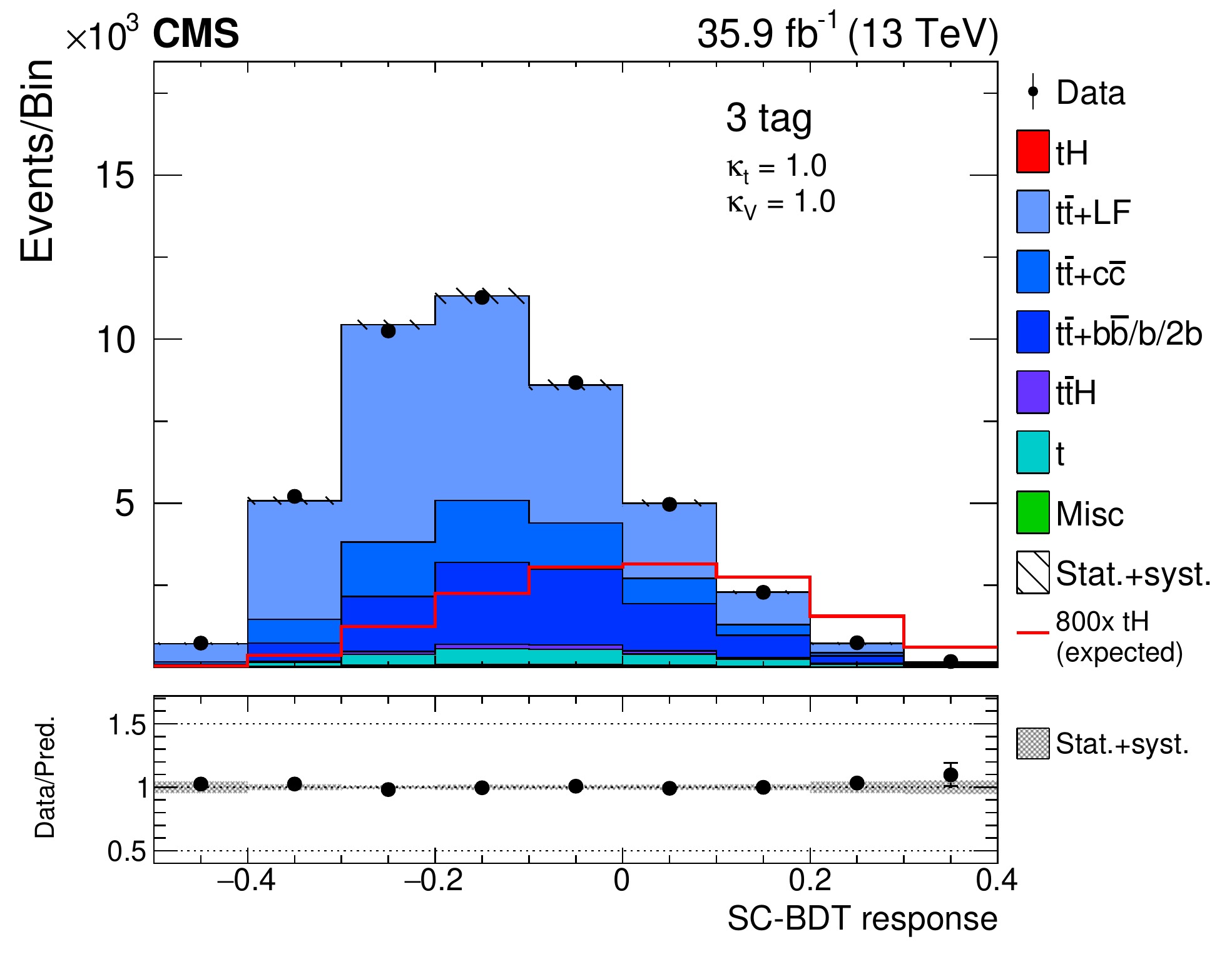}
\includegraphics[width=0.48\textwidth]{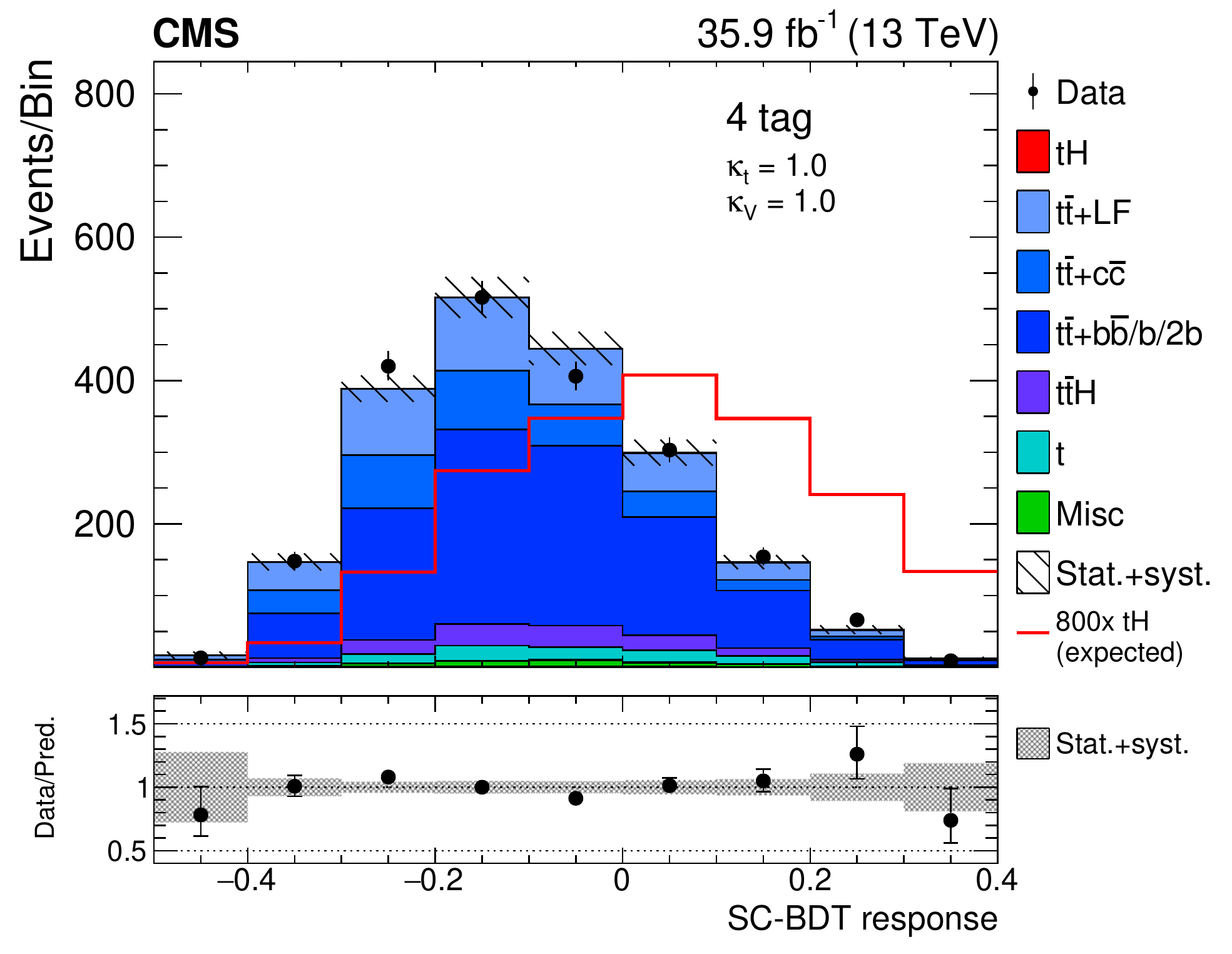}
\end{center}
\caption{Post-fit classifier outputs of the signal classification BDT as used in the maximum likelihood fit for the 3 tag channel (left) and the 4 tag channel (right). In the box below each distribution, the ratio of the observed and predicted event yields is shown. The shape of the \tH\ signal is indicated with 800 times the SM.
\label{fig:bb_postfit}}
\end{figure*}

\begin{figure*}[!htb]
\begin{center}
\includegraphics[width=0.48\textwidth]{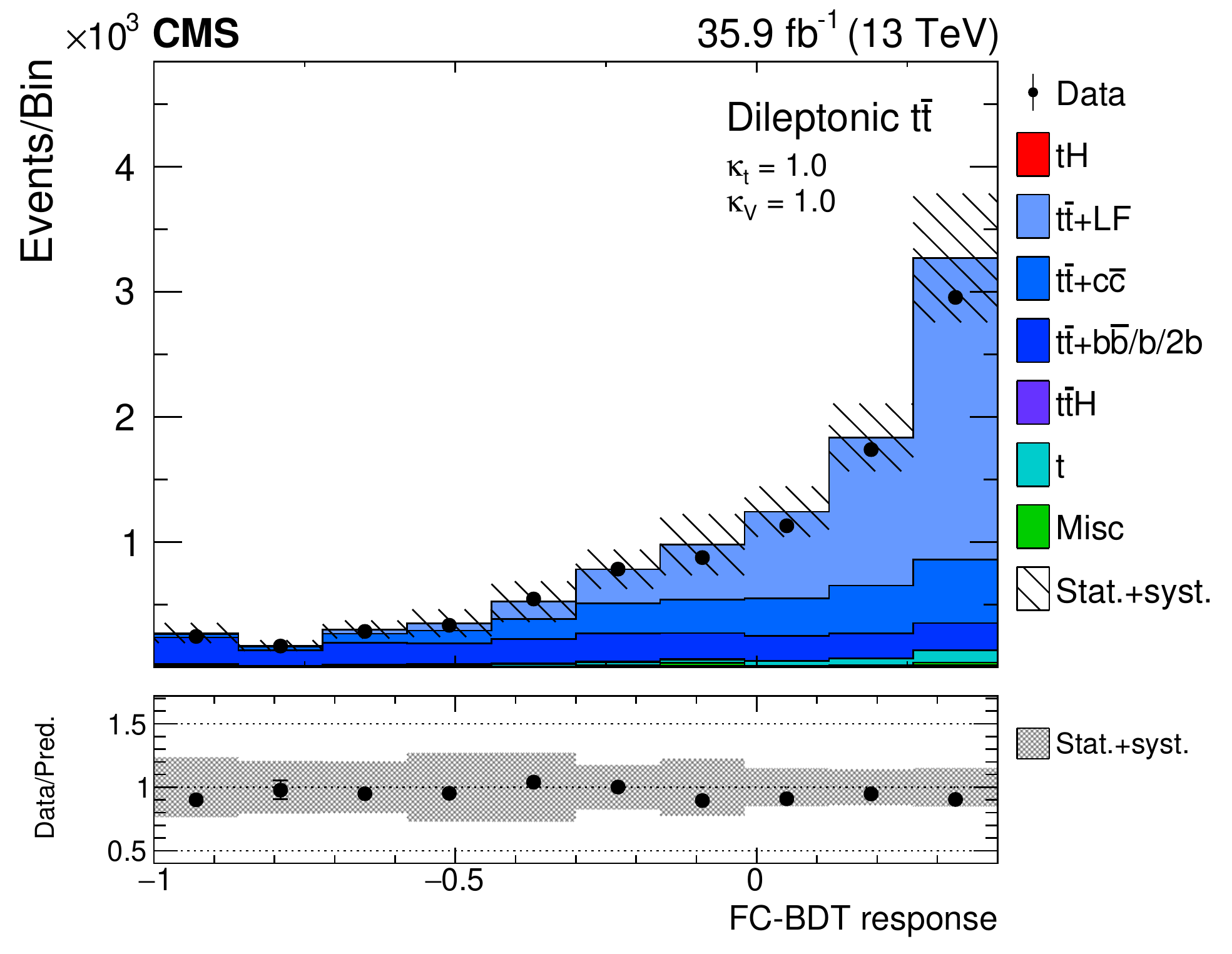}
\includegraphics[width=0.48\textwidth]{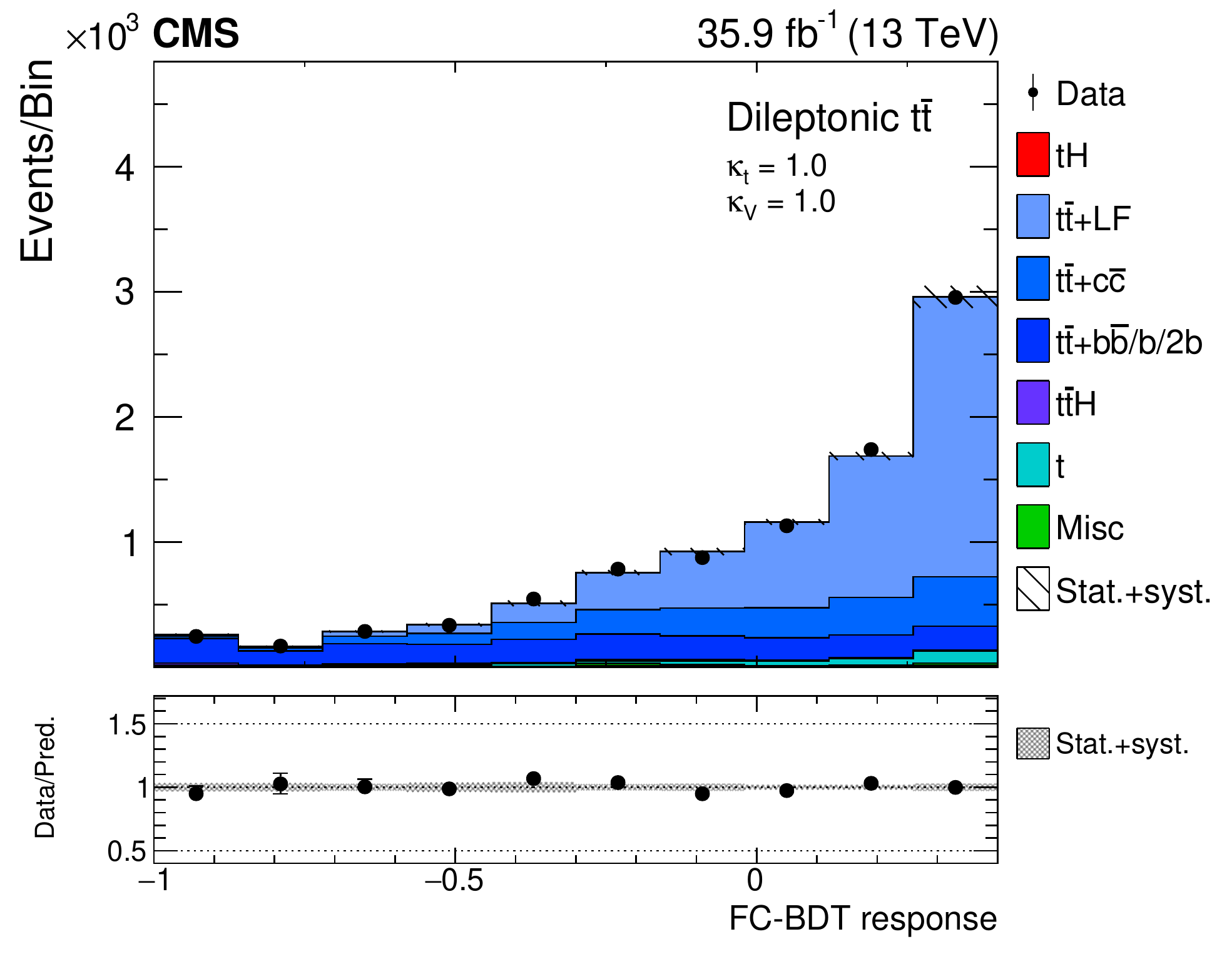}
\end{center}
\caption{Pre- (left) and post-fit (right) classifier outputs of the flavor classification BDT for the dilepton selection. In the box below each distribution, the ratio of the observed and predicted event yields is shown.
\label{fig:bb_dilepton}}
\end{figure*}

To determine the signal yield, the output distributions of the SC-BDT in the three and four \cPqb\ tag regions are fitted simultaneously with the output of the FC-BDT in the dilepton region.
The SC-BDT output distributions before the fit are shown in Fig.~\ref{fig:bb_prefit} and the result of the fit is shown in Fig.~\ref{fig:bb_postfit}.
The pre- and post-fit distributions of the FC-BDT are shown in Fig.~\ref{fig:bb_dilepton}.

\subsection{Systematic uncertainties}\label{ssec:bbsystematics}
Many systematic uncertainties affect the result of the analysis, arising both from experimental and theoretical sources.
All uncertainties are parametrized as nuisance parameters in the statistical inference performed in the final analysis step described in Section~\ref{sec:results}.

The uncertainty in the signal normalization due to the choice of factorization and renormalization scales is evaluated by changing their values to double and half of the nominal values.
A rate uncertainty of around $5\%$ is assigned to each process to account for the choice of PDFs, since shape variations are found to be negligible.
Furthermore, for each \tthf\ category, an individual $50\%$ rate uncertainty is assigned, since the modeling of these components is theoretically difficult and cross section measurements are affected by large systematic uncertainties~\cite{Khachatryan:2015mva,Sirunyan:2017snr}.

The observed top quark \pt\ spectrum in \ttbar\ events is found to be softer than the theoretical prediction~\cite{Khachatryan:2015oqa}.
A systematic uncertainty for this effect is derived by applying event-by-event weights that correct the disagreement.

Efficiency corrections for the selection of isolated leptons by the trigger and quality requirements are evaluated with a tag-and-probe method.
Uncertainties in correcting the distribution of PV interactions are accounted by varying the total inelastic cross section by $4.6\%$~\cite{Sirunyan:2018nqx}.
The corrections applied to the jet energy scale and resolution are varied within their given uncertainties and the migration between different categories is used to determine the effect.
In addition, the contribution to \ptmiss\ of unclustered particles is varied within the resolution of each particle~\cite{Khachatryan:2014gga}.
The \cPqb\ tagging efficiencies for jets are measured in QCD multijet and \ttbar\ enriched samples and varied within their uncertainties~\cite{Sirunyan:2017ezt}.

As for the multilepton channel, an uncertainty of $2.5\%$ is assigned to the integrated luminosity~\cite{LUM-17-001} and affects the normalization of all processes.

The dominant systematic uncertainties are those related to the factorization and renormalization scales, as well as the uncertainties in the overall normalization of the \tthf\ processes and the jet energy corrections.

\section{Reinterpretation of the \texorpdfstring{$\PH\to\gamgam$}{H to gamma gamma} measurement}\label{sec:aa}
The standard model \tHq\ and \tHW\ signal processes with $\PH\to\Pgg\Pgg$ were included in previous measurements of the Higgs boson properties in the inclusive diphoton final state~\cite{Sirunyan:2018ouh}.
Events with two prompt high-\pt\ photons were divided into different event categories, each enriched with a particular production mechanism of the Higgs boson.
The \tHq\ and \tHW\ processes contribute mostly to the ``\ttH\ hadronic'', and ``\ttH\ leptonic'' categories as defined in Ref.~\cite{Sirunyan:2018ouh}, which target the \ttH\ process for fully hadronic top quark decays and for single-lepton or dilepton decay modes, respectively.
Events in the \ttH\ leptonic category are selected to have at least one lepton well separated from the photons, and well reconstructed, as well as at least two jets of which at least one passes the medium \cPqb\ tagging requirement.
The \ttH\ hadronic category is defined as events with no identified leptons and at least three jets, of which at least one is \cPqb\ tagged with the loose working point.

The signal is modeled with a sum of Gaussian functions describing the diphoton invariant mass ($m_{\Pgg\Pgg}$) shape derived from simulation.
The background contribution is determined from the data without reliance on simulated events, using the discrete profiling method~\cite{Dauncey:2014xga,Khachatryan:2014ira,Aad:2015zhl}.
Different classes of models describing the falling $m_{\Pgg\Pgg}$ distribution in the background processes are used as input to the method.
Sources of systematic uncertainties affecting the signal model and leading to migrations of signal events among the categories are considered.

\begin{figure}[htb]
  \centering
  \includegraphics[width=\cmsFigWidth]{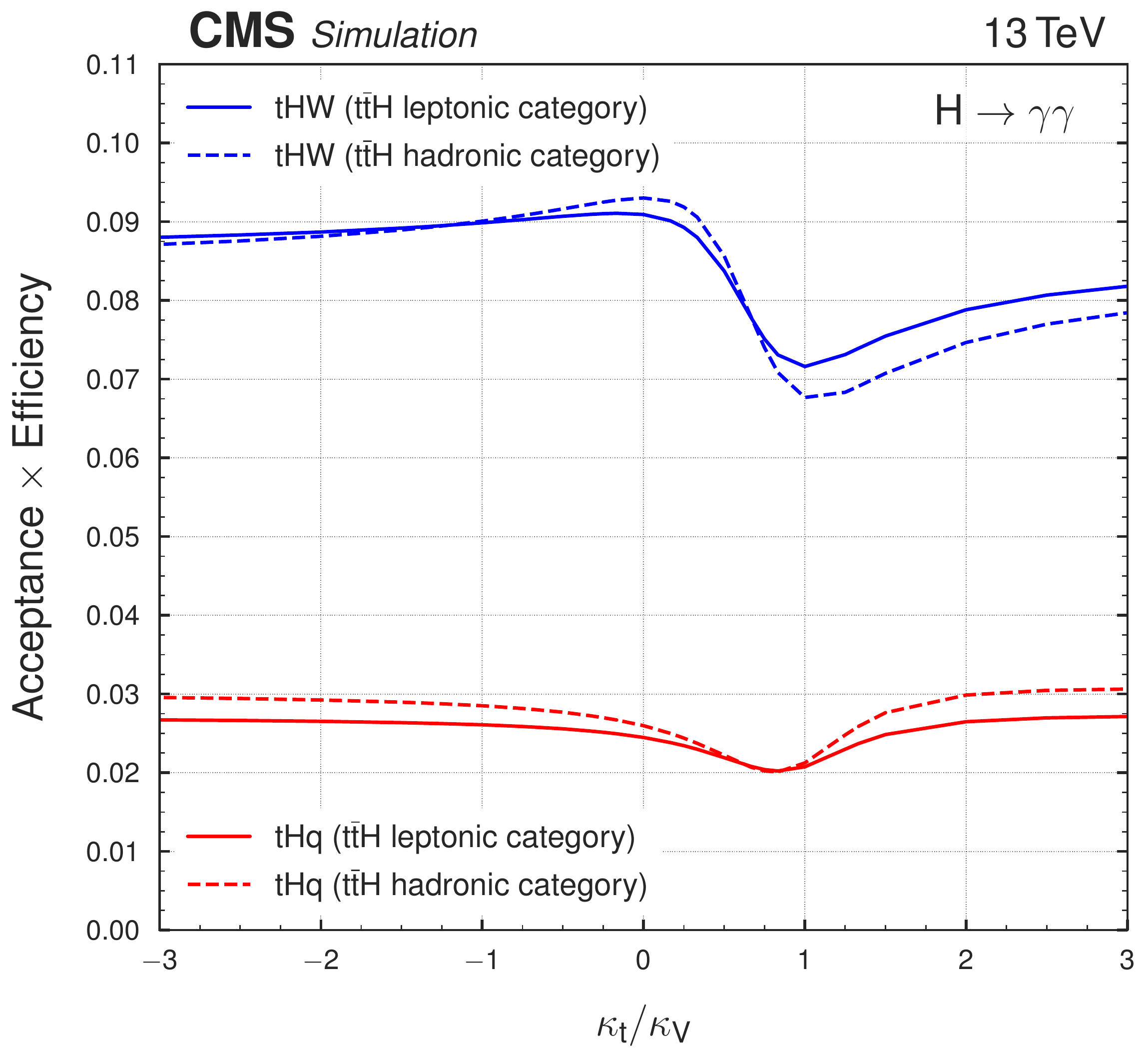}
  \caption{Acceptance and selection efficiency for the \tHq\ (red) and \tHW\ (blue) signal processes as a function of $\kappat/\kappaV$, for the \ttH\ leptonic (solid lines) and \ttH hadronic categories (dashed lines) of the $\PH\to\gamgam$ measurement.}
  \label{fig:effaa}
\end{figure}

The inputs to Ref.~\cite{Sirunyan:2018ouh} from the \ttH\ categories are used here in a combination with the multilepton and single-lepton + \bbbar\ channels to put constraints on the coupling modifier \kappat\ and on the production cross section of \tH\ events.
The coupling modifiers \kappat\ and \kappaV\ affect both the \tH\ and \ttH\ production cross sections, as well as the Higgs boson decay branching fraction into two photons through the interference of bosonic and fermionic loops.
Changes in the kinematic properties of the \tH\ signal arising from the modified couplings are taken into account by considering their effect on the signal acceptance and selection efficiency.
Figure~\ref{fig:effaa} shows the modified \tHq\ and \tHW\ selection efficiencies including acceptances for the two relevant categories of the $\PH\to\gamgam$ measurement as a function of the ratio of coupling modifiers $\kappat/\kappaV$.
The signal diphoton mass shape is found to be independent of $\kappat/\kappaV$.

The dependence of the signal acceptance and efficiency on $\kappat/\kappaV$ is implemented in the same statistical framework as that of Ref.~\cite{Sirunyan:2018ouh}, modifying the signal only in the \ttH\ categories.

\section{Results and interpretation}\label{sec:results}
The different discriminator output distributions in the multilepton and single-lepton + \bbbar\ channels and the $\gamgam$ invariant mass distributions in the diphoton channel are compared to the data in a combined maximum likelihood fit for various assumptions on the signal kinematics and normalizations, and are used to derive constraints on the signal yields.

The event selections in the different channels are mutually exclusive, therefore allowing a straightforward combination.
Common systematic uncertainties such as the integrated luminosity normalization, the \cPqb\ tagging uncertainties, and the theoretical uncertainties related to the signal modeling are taken to be fully correlated among different channels.

A profile likelihood scan is performed as a function of the coupling modifier \kappat, which affects the production cross sections of the three signal components \tHq, \tHW, and \ttH, as well as the Higgs boson branching fractions.
Effects on Higgs boson decays via fermion and boson loops to \gamgam, $\PZ\gamma$, and gluon-gluon final states also affect the branching fractions in other channels.
Furthermore, the kinematic properties of the two \tH\ processes and thereby the shape of the classifier outputs entering the fit depend on the value of $\kappat$.
Systematic uncertainties are included in the form of nuisance parameters in the fit and are treated via the frequentist paradigm, as described in Refs.~\cite{ATL-PHYS-PUB-2011-011,HIG-11-032}.
Uncertainties affecting the normalization are constrained either by $\Gamma$-function distributions, if they are statistical in origin, or by log-normal probability density functions.
Systematic uncertainties that affect both the normalization and shape in the discriminating observables are included in the fit using the technique detailed in Ref.~\cite{Conway:2011in}, and represented by Gaussian probability density functions.

Table~\ref{tab:systematics} shows the impact of the most important groups of nuisances parameters on the \tHttH\ signal yield.
Pre-fit systematic uncertainties of the same groups are shown for comparison.

\begin{table}[htb]
  \topcaption{Summary of the main sources of systematic uncertainty. $\Delta\mu/\mu$ corresponds to the relative change in \tHttH\ signal yield induced by varying the systematic source within its associated uncertainty.\label{tab:systematics}}
  \begin{center}
    \begin{scotch}{lcc}
    Source                          & Uncertainty [\%]   & $\Delta\mu/\mu$ [\%]\\
    \hline
    \Pe, \Pgm\ selection efficiency & 2--4               & 17  \\
    $\cPqb$ tagging efficiency      & 2--15              & 6   \\
    Jet energy calibration          & 2--15              & 3   \\
    Forward jet modeling            & 10--35             & 3   \\
    Integrated luminosity           & 2.5                & 10  \\
    Reducible background estimate   & 10--40             & 14  \\
    Theoretical sources             & $\approx$10        & 14  \\
    \tthf\ normalization            & $\approx$50        & 7   \\
    PDFs                            & 2--6               & 8   \\
    \end{scotch}
  \end{center}
\end{table}

\begin{figure}[htb]
  \centering
  \includegraphics[width=\cmsFigWidth]{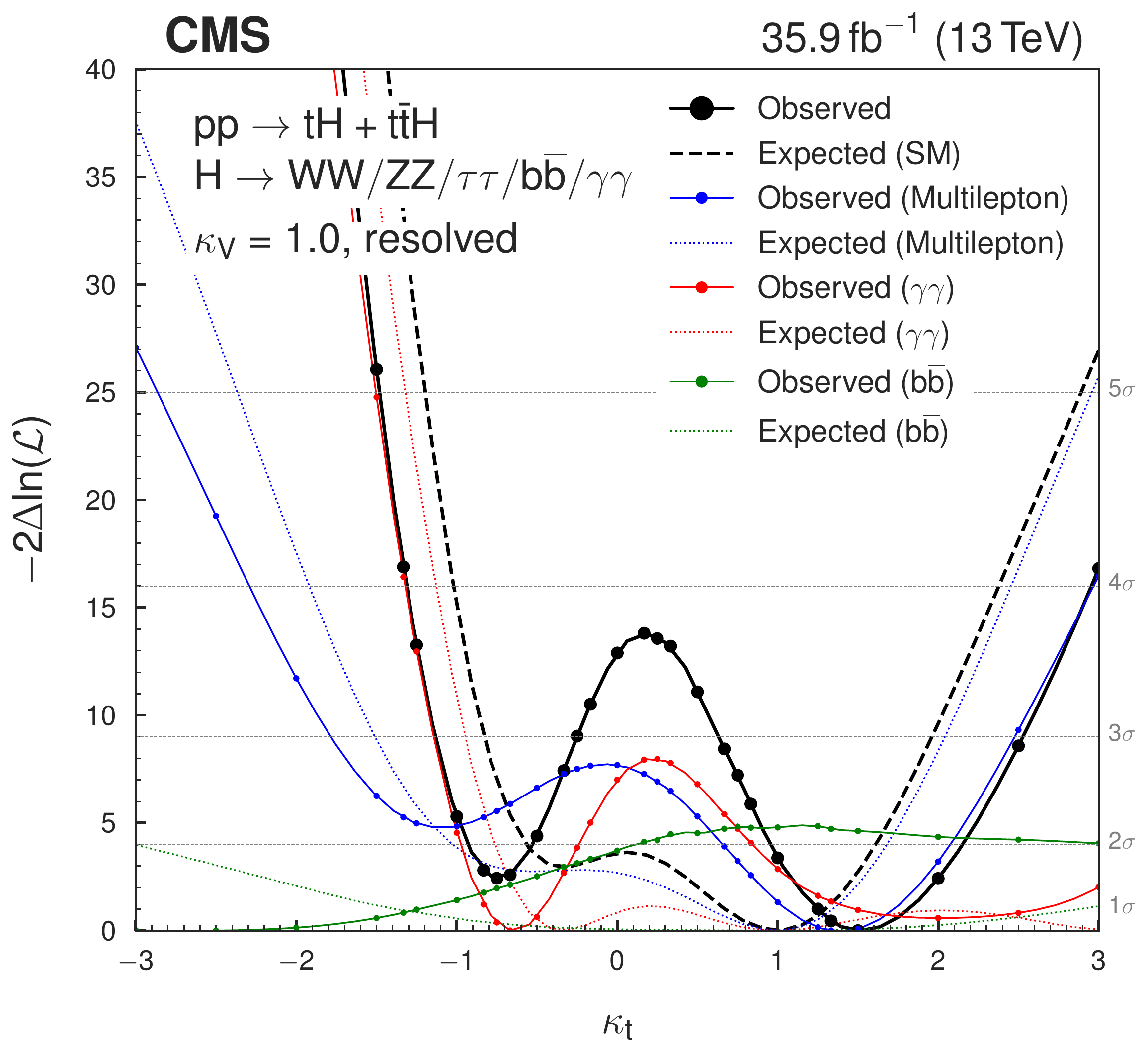}
  \caption{Scan of $-2\Delta\ln{(\mathcal{L})}$ versus \kappat\ for the data (black line) and the individual channels (blue, red, and green), compared to Asimov data sets corresponding to the SM expectations (dashed lines).}
  \label{fig:nll_scan}
\end{figure}

To derive constraints on \kappat\ for a fixed value of $\kappaV=1.0$, a scan of the likelihood ratio $\mathcal{L}(\kappat)/\mathcal{L}(\hat{\kappat})$ is performed, where $\hat{\kappat}$ is the best fit value of \kappat.
Figure~\ref{fig:nll_scan} shows the negative of twice the logarithm of this likelihood ratio ($-2\Delta\ln{(\mathcal{L})}$), for scans on the data, and for an Asimov data set~\cite{Cowan2011} with SM expectations for \ttH\ and \tH.
On this scale, a 95\% confidence interval covers values below 3.84, while standard deviations are at values of 1, 4, 9, 16, \etc
The expected performance for an SM-like signal is to favor a value of $\kappat=1.0$ over one of $\kappat=-1.0$ by more than four standard deviations, and to exclude values outside of about $-0.5$ and $1.6$ at 95\% confidence level (\CL).
In the combined scan, the data slightly favor a positive value of \kappat\ over a negative one, by about 1.5 standard deviations, while excluding values outside the ranges of about $[-0.9, -0.5]$ and $[1.0, 2.1]$ at 95\% \CLnp.
The sensitivity is driven by the \gamgam\ channel at negative values of the coupling modifiers and by the multilepton channels at positive values.

An excess of observed over expected events is seen both in the multilepton and \gamgam\ channels, with a combined significance of about two standard deviations.
Consequently, floating a signal strength modifier (defined as the ratio of the fitted signal cross section to the SM expectation) of a combined \tHttH\ signal yields a best fit value of $2.00\pm0.53$ under the SM hypothesis.
These results are in agreement with those from the dedicated \ttH\ searches~\cite{cms_tthobs}, as expected, since they share a large fraction of events with the data set used here.

\begin{figure}[htb]
  \centering
  \includegraphics[width=\cmsFigWidth]{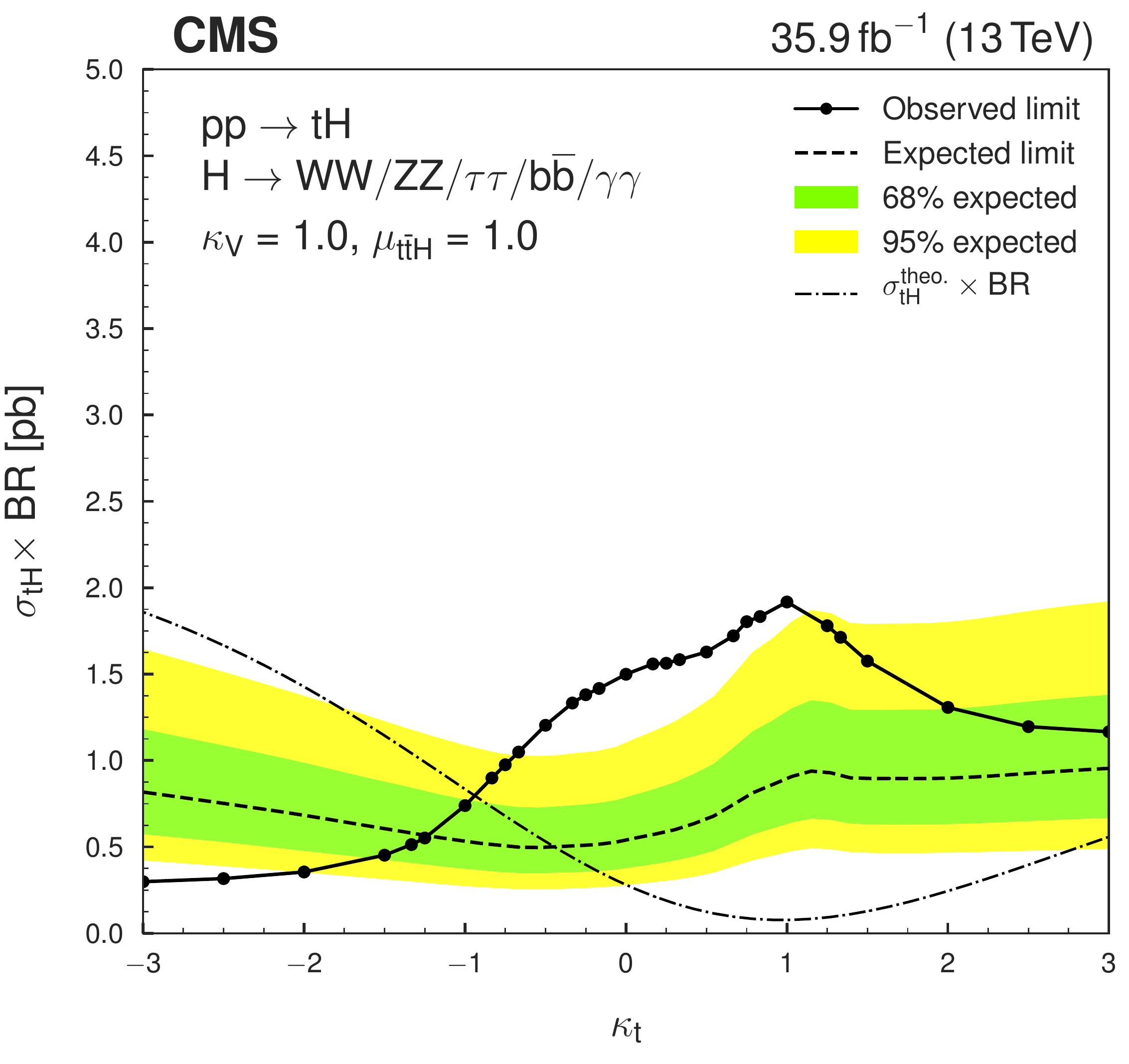}
  \caption{Observed and expected 95\% \CL upper limit on the \tH\ cross section times combined $\PH\to\PW\PW^*+\tautau+\PZ\PZ^*+\bbbar+\gamgam$ branching fraction for different values of the coupling ratio \kappat. The expected limit is calculated on a background-only data set, \ie, without \tH\ contribution, but including a \kappat-dependent contribution from \ttH. The \ttH\ normalization is kept fixed in the fit, while the \tH\ signal strength is allowed to float.
  }
  \label{fig:limits_smexp}
\end{figure}

To establish limits on \tH\ production, a different signal strength parameter is introduced for the combination of \tHq\ and \tHW, not including \ttH.
A maximum likelihood fit for this signal strength is then performed based on the profile likelihood test statistic~\cite{ATL-PHYS-PUB-2011-011,HIG-11-032} at fixed points of \kappat.
Upper limits on the signal strength are then derived using the \CLs method~\cite{Junk:1999kv,Read:2002hq} and using asymptotic formulae for the distribution of the test statistic~\cite{Cowan2011}.
They are multiplied by the \kappat-dependent \tH\ production cross section times the combined Higgs boson branching fractions to $\PW\PW^*+\tautau+\PZ\PZ^*+\bbbar+\gamgam$ and are shown in Fig.~\ref{fig:limits_smexp}.
Limits for the SM and for a scenario with $\kappat=-1.0$ for the individual channels are shown in Table~\ref{tab:limits}.
The \ttH\ contribution is kept fixed to its \kappat-dependent expectation.
The fiducial cross section for SM-like \tH\ production is limited to about 1.9\unit{pb}, with an expected limit of 0.9\unit{pb}, corresponding, respectively, to about 25 and 12 times the expected cross section times branching fraction in the combination of the channels explored.
The visible discrepancy between observed and expected limits around $\kappat=0.0$ is caused by the fact that the predicted \ttH\ cross section vanishes in that scenario while the data favor even larger than expected yields for \ttH.

\begin{table}[htb]
  \topcaption{Expected and observed 95\% \CL upper limits on the \tH\ production cross section times $\PH\to\PW\PW^*+\tautau+\PZ\PZ^*+\bbbar+\gamgam$ branching fraction for a scenario of inverted couplings ($\kappat=-1.0$, top rows), vanishing top quark Yukawa coupling ($\kappat=0.0$, middle rows), and for an SM-like signal ($\kappat=1.0$, bottom rows), in pb. The expected limit is calculated on a background-only data set, \ie, without \tH\ contribution, but including a \kappat-dependent contribution from the \ttH production. The \ttH\ normalization is kept fixed in the fit, while the \tH\ signal strength is allowed to float. Limits can be compared to the expected product of \tH\ cross sections and branching fractions of 0.83, 0.28, and 0.077\unit{pb} for the inverted top quark Yukawa coupling, the $\kappat=0$ scenario, and for the SM, respectively.\label{tab:limits}}
  \begin{center}
    \begin{scotch}{llcc}
      Scenario  & Channel         & Observed & Expected         \\
   \multirow{4}{*}{$\kappat=-1$}
                & \bbbar\                  & $4.98\unit{pb}$ & $2.52\,^{+1.29}_{-0.81}\unit{pb}$ \\
                & \gamgam\                 & $0.84\unit{pb}$ & $0.88\,^{+0.46}_{-0.28}\unit{pb}$ \\
                & $\mupmup+\epmup+\threel$ & $0.85\unit{pb}$ & $0.77\,^{+0.36}_{-0.24}\unit{pb}$ \\
                & Combined                 & $0.74\unit{pb}$ & $0.53\,^{+0.24}_{-0.16}\unit{pb}$ \\
                [\cmsTabSkip]
   \multirow{4}{*}{\shortstack[l]{$\kappat=0$ \\ ($\ttH=0$)}}
                & \bbbar\                  & $5.18\unit{pb}$ & $2.60\,^{+1.35}_{-0.84}\unit{pb}$ \\
                & \gamgam\                 & $2.63\unit{pb}$ & $0.96\,^{+0.50}_{-0.31}\unit{pb}$ \\
                & $\mupmup+\epmup+\threel$ & $0.83\unit{pb}$ & $0.76\,^{+0.36}_{-0.23}\unit{pb}$ \\
                & Combined                 & $1.50\unit{pb}$ & $0.54\,^{+0.25}_{-0.16}\unit{pb}$ \\
                [\cmsTabSkip]
   \multirow{4}{*}{\shortstack[l]{$\kappat=1$ \\ (SM-like)}}
                & \bbbar\                  & $6.88\unit{pb}$ & $3.19\,^{+1.64}_{-1.02}\unit{pb}$ \\
                & \gamgam\                 & $3.68\unit{pb}$ & $2.03\,^{+1.05}_{-0.67}\unit{pb}$ \\
                & $\mupmup+\epmup+\threel$ & $1.36\unit{pb}$ & $1.18\,^{+0.53}_{-0.35}\unit{pb}$ \\
                & Combined                 & $1.94\unit{pb}$ & $0.92\,^{+0.40}_{-0.27}\unit{pb}$ \\
    \end{scotch}
  \end{center}
\end{table}

\section{Summary}\label{sec:conclusion}
Events from proton-proton collisions at $\sqrt{s}=13\TeV$ compatible with the production of a Higgs boson (\PH) in association with a single top quark (\cPqt) have been studied to derive constraints on the magnitude and relative sign of Higgs boson couplings to top quarks and vector bosons.
Dedicated analyses of multilepton final states and final states with single leptons and a pair of bottom quarks are combined with a reinterpretation of a measurement of Higgs bosons decaying to two photons for the final result.
For standard model-like Higgs boson couplings to vector bosons, the data favor a positive value of the top quark Yukawa coupling, \yt, by 1.5 standard deviations and exclude values outside the ranges of about $[-0.9, -0.5]$ and $[1.0, 2.1]$ times $\yt^\mathrm{SM}$ at the 95\% confidence level.
An excess of events compared with expected backgrounds is observed, but it is still compatible with the standard model expectation for \tHttH\ production.

\begin{acknowledgments}
We congratulate our colleagues in the CERN accelerator departments for the excellent performance of the LHC and thank the technical and administrative staffs at CERN and at other CMS institutes for their contributions to the success of the CMS effort. In addition, we gratefully acknowledge the computing centers and personnel of the Worldwide LHC Computing Grid for delivering so effectively the computing infrastructure essential to our analyses. Finally, we acknowledge the enduring support for the construction and operation of the LHC and the CMS detector provided by the following funding agencies: BMBWF and FWF (Austria); FNRS and FWO (Belgium); CNPq, CAPES, FAPERJ, FAPERGS, and FAPESP (Brazil); MES (Bulgaria); CERN; CAS, MoST, and NSFC (China); COLCIENCIAS (Colombia); MSES and CSF (Croatia); RPF (Cyprus); SENESCYT (Ecuador); MoER, ERC IUT, and ERDF (Estonia); Academy of Finland, MEC, and HIP (Finland); CEA and CNRS/IN2P3 (France); BMBF, DFG, and HGF (Germany); GSRT (Greece); NKFIA (Hungary); DAE and DST (India); IPM (Iran); SFI (Ireland); INFN (Italy); MSIP and NRF (Republic of Korea); MES (Latvia); LAS (Lithuania); MOE and UM (Malaysia); BUAP, CINVESTAV, CONACYT, LNS, SEP, and UASLP-FAI (Mexico); MOS (Montenegro); MBIE (New Zealand); PAEC (Pakistan); MSHE and NSC (Poland); FCT (Portugal); JINR (Dubna); MON, RosAtom, RAS, RFBR, and NRC KI (Russia); MESTD (Serbia); SEIDI, CPAN, PCTI, and FEDER (Spain); MOSTR (Sri Lanka); Swiss Funding Agencies (Switzerland); MST (Taipei); ThEPCenter, IPST, STAR, and NSTDA (Thailand); TUBITAK and TAEK (Turkey); NASU and SFFR (Ukraine); STFC (United Kingdom); DOE and NSF (USA).

\hyphenation{Rachada-pisek} Individuals have received support from the Marie-Curie program and the European Research Council and Horizon 2020 Grant, contract No. 675440 (European Union); the Leventis Foundation; the A. P. Sloan Foundation; the Alexander von Humboldt Foundation; the Belgian Federal Science Policy Office; the Fonds pour la Formation \`a la Recherche dans l'Industrie et dans l'Agriculture (FRIA-Belgium); the Agentschap voor Innovatie door Wetenschap en Technologie (IWT-Belgium); the F.R.S.-FNRS and FWO (Belgium) under the ``Excellence of Science - EOS" - be.h project n. 30820817; the Ministry of Education, Youth and Sports (MEYS) of the Czech Republic; the Lend\"ulet (``Momentum") Program and the J\'anos Bolyai Research Scholarship of the Hungarian Academy of Sciences, the New National Excellence Program \'UNKP, the NKFIA research grants 123842, 123959, 124845, 124850 and 125105 (Hungary); the Council of Science and Industrial Research, India; the HOMING PLUS program of the Foundation for Polish Science, cofinanced from European Union, Regional Development Fund, the Mobility Plus program of the Ministry of Science and Higher Education, the National Science Center (Poland), contracts Harmonia 2014/14/M/ST2/00428, Opus 2014/13/B/ST2/02543, 2014/15/B/ST2/03998, and 2015/19/B/ST2/02861, Sonata-bis 2012/07/E/ST2/01406; the National Priorities Research Program by Qatar National Research Fund; the Programa Estatal de Fomento de la Investigaci{\'o}n Cient{\'i}fica y T{\'e}cnica de Excelencia Mar\'{\i}a de Maeztu, grant MDM-2015-0509 and the Programa Severo Ochoa del Principado de Asturias; the Thalis and Aristeia programs cofinanced by EU-ESF and the Greek NSRF; the Rachadapisek Sompot Fund for Postdoctoral Fellowship, Chulalongkorn University and the Chulalongkorn Academic into Its 2nd Century Project Advancement Project (Thailand); the Welch Foundation, contract C-1845; and the Weston Havens Foundation (USA). \end{acknowledgments}

\bibliography{auto_generated}
\cleardoublepage \appendix\section{The CMS Collaboration \label{app:collab}}\begin{sloppypar}\hyphenpenalty=5000\widowpenalty=500\clubpenalty=5000\vskip\cmsinstskip
\textbf{Yerevan Physics Institute, Yerevan, Armenia}\\*[0pt]
A.M.~Sirunyan, A.~Tumasyan
\vskip\cmsinstskip
\textbf{Institut f\"{u}r Hochenergiephysik, Wien, Austria}\\*[0pt]
W.~Adam, F.~Ambrogi, E.~Asilar, T.~Bergauer, J.~Brandstetter, M.~Dragicevic, J.~Er\"{o}, A.~Escalante~Del~Valle, M.~Flechl, R.~Fr\"{u}hwirth\cmsAuthorMark{1}, V.M.~Ghete, J.~Hrubec, M.~Jeitler\cmsAuthorMark{1}, N.~Krammer, I.~Kr\"{a}tschmer, D.~Liko, T.~Madlener, I.~Mikulec, N.~Rad, H.~Rohringer, J.~Schieck\cmsAuthorMark{1}, R.~Sch\"{o}fbeck, M.~Spanring, D.~Spitzbart, A.~Taurok, W.~Waltenberger, J.~Wittmann, C.-E.~Wulz\cmsAuthorMark{1}, M.~Zarucki
% ...
\vskip\cmsinstskip
% ...
1:  Also at Vienna University of Technology, Vienna, Austria\\
% ...
\end{sloppypar}
\end{document}